\documentclass[12pt]{article}
\usepackage{graphicx}
\usepackage{lscape}
\usepackage{epsfig}
\usepackage{citesort}
\usepackage{amssymb}
\usepackage{amsmath}
\usepackage{multirow}
\usepackage[table]{xcolor}
\usepackage{colortbl}
\definecolor{lightgray}{gray}{0.9}

\newcommand{\be}{\begin{equation}}
\newcommand{\ee}{\end{equation}}
\newcommand{\bea}{\begin{eqnarray}}
\newcommand{\eea}{\end{eqnarray}}
\newcommand{\nn}{\nonumber}

\def\R1{\varepsilon_1}
\def\E8{\varepsilon_8}

\def\ga{\gamma}

\def\lb{\Lambda_b}

\def\s1{\hat s}
\def\ds{\displaystyle}

\newcommand{\bd}{\begin{displaymath}}
\newcommand{\ed}{\end{displaymath}}

\def\R1{\varepsilon_1}
\def\E8{\varepsilon_8}

\def\ga{\gamma}

\def\ds{\displaystyle}
\def\beq{\begin{equation}}
\def\eeq{\end{equation}}
\def\bea{\begin{eqnarray}}
\def\eea{\end{eqnarray}}
\def\beeq{\begin{eqnarray}}
\def\eeeq{\end{eqnarray}}

\def\vel{\left|}
\def\ver{\right|}
\def\nnb{\nonumber}
\def\ga{\left(}
\def\dr{\right)}

\def\rar{\rightarrow}
\def\nnb{\nonumber}

\def\ba{\begin{array}}
\def\ea{\end{array}}

\def\xis0{{\Xi^{*0}}}

\def\g5{\gamma_5}

\def\es{\!\!\! &=& \!\!\!}

\def\ar{&+& \!\!\!}
\def\ek{&-& \!\!\!}

\begin{document}
\title{
         {\Large
                 {\bf Comparative analysis of the semileptonic $\Lambda_b
\rightarrow \Lambda \ell^+ \ell^-$ transition
in SM and different SUSY scenarios using form factors from full QCD
                 }
         }
      }
      
\author{\vspace{1cm}\\
{\small  K. Azizi$^1$ \thanks {e-mail: kazizi@dogus.edu.tr}\,\,, S. Kartal$^2$ \thanks
{e-mail: sehban@istanbul.edu.tr}\,\,, A. T. Olgun$^2$ \thanks
{e-mail: a.t.olgun@gmail.com}\,\,, Z. Tavuko\u glu$^2$ \thanks
{e-mail: z.tavukoglu@gmail.com}}  \\
{\small $^1$ Department of Physics, Do\u gu\c s University,
Ac{\i}badem-Kad{\i}k\"oy, 34722 \.{I}stanbul, Turkey}\\
{\small $^2$ Department of Physics, \.{I}stanbul University,
Vezneciler, 34134 \.{I}stanbul, Turkey}\\
}

\date{}

\begin{titlepage}
\maketitle
\thispagestyle{empty}
\begin{abstract}
We work out the semileptonic $\Lambda _b\rightarrow \Lambda \ell^+ \ell^-$ transition in  standard   as well as  different supersymmetric models. In particular, considering the parametrization of the matrix elements entered the low energy effective Hamiltonian in terms of form factors in full QCD, we calculate the amplitude and differential decay rate responsible for this decay channel in supersymmetric models. We then use the form factors calculated via light cone QCD sum rules in full theory to analyze the differential branching ratio and  lepton forward-backward asymmetry of this decay channel  in different supersymmetric models and compare the obtained results with those of the standard model. We also discuss how the results of different supersymmetric models deviate from the standard model predictions and which SUSY scenarios are favored.
 
\end{abstract}

~~~PACS number(s): 12.60.-i, 12.60.Jv, 13.30.-a, 13.30.Ce, 14.20.Mr 
\end{titlepage}



\section{Introduction}
Recently, there has been an important progress on the course of searching for Higgs boson as a missing 
ingredient of the standard model (SM). 
The ATLAS and CMS Collaborations at CERN reported their observation on a Higgs-like particle with a statistical significance
of $5\sigma$ \cite{CMSATLAS}. Now, it is searched whether this Higgs-like boson is the standard or non-standard Higgs
particle.  The  supersymmetry (SUSY) has been the most popular paradigm for new physics (NP) scenarios in the last decades. The recent progresses have stimulated
the  theoretical works dedicated to the study of how a relatively heavy Higgs constrains the  parameters of SUSY (for a discussion see for instance \cite{Hirsch}). On the other hand, with these  developments,
 we hope that we will have an experimental  progress in searching for SUSY particles both directly by increasing the center of mass energy and indirectly by studying the flavor changing neutral current (FCNC) transitions.


In the present work, we theoretically analyze the semileptonic FCNC decay of the $\Lambda _b \rightarrow \Lambda \ell^+ \ell^-$ in existing related different supersymmetric models. In principle, the SUSY particles
 can contribute to such loop level transitions. Hence, we look for the effect of superparticles in this channel via calculating some related observables like differential branching ratio and 
lepton forward-backward asymmetry (FBA). Due to the specific features, there are different SUSY scenarios such as SUSY I, SUSY II, SUSY III and SUSY SO(10) \cite{Aslam,Aslam2,SUSY,Wil.coef}. 
In these models, the Wilson coefficients receive contributions from neutral Higgs bosons (NHBs) that are proportional to $ tan^{3}\beta $, where $ tan\beta $ has been defined as
 the ratio of the vacuum expectation values of two neutral Higgs bosons ($ h^{0}, A^{0}$). According to the $ tan\beta $ and an extra parameter $ \mu $ with dimension of mass corresponding to mass 
term mixing of
 two Higgs doublets, the different SUSY models are categorized. In SUSY I, the $ \mu $ takes negative value, some of the Wilson coefficients change their signs and the contributions of
 NHBs have been neglected. In SUSY II, the $ tan\beta $ takes large value while masses of the superparticles are small in order of a few hundred GeV. In SUSY III, the $ tan\beta $ is 
large and the masses of the superparticles are relatively large up to 450 GeV or more.  In SUSY SO(10) model, the imaginary parts of the Wilson 
coefficients are large and the NHBs contributions are considered. 

In the last year, the CDF Collaboration at Fermilab \cite{CDF-lambdab} has  reported the first observation on the baryonic FCNC transition of $\Lambda_b^0 \rightarrow \Lambda \mu ^+ \mu ^-$ 
with 24 signal events and statistical of 5.8 deviations. They have measured a branching ratio of $[1.73\pm 0.42(stat)\pm 0.55(syst)]\times 10^{-6}$. This decay channel is in 
the focus of different experiments like LHCb at CERN \cite{LHCb}. Hence, theoretical and phenomenological predictions on the observables defining this channel can help us in 
the course of searching indirectly for SUSY particles in this stage. Comparison of different theoretical results with experimental data may help us get useful informations about
 the existence of the SUSY particles. Note that the rare  $\Lambda _b \rightarrow \Lambda \ell^+ \ell^-$ transition was analyzed in the same frameworks in \cite{Aslam} using only two form factors calculated 
via heavy quark effective theory (HQET). In this work, we generalize those calculations to include all form factors in full theory.

In the next section, introducing the effective Hamiltonian both in the SM and SUSY models, we calculate the amplitude of the decay under consideration in terms of twelve form factors enrolled 
to the transition matrix elements. In  section 3, we calculate the formula for the differential decay rate in SUSY and numerically analyze it together with the branching ratio and lepton FBA. We also compare the obtained results on the considered physical quantities in different SUSY models with those obtained from the SM. The last section encompasses our concluding remarks.


\section{The effective Hamiltonian and transition matrix elements }

In the SM, the $\Lambda _b \rightarrow \Lambda \ell^+ \ell^-$ transition goes with the $b \rar s \ell^+ \ell^- $ at quark level whose effective Hamiltonian  is given by 
\bea \label{Heff} {\cal H}^{eff} &=& {G_F \alpha_{em} V_{tb}
V_{ts}^\ast \over 2\sqrt{2} \pi} \Bigg[ C^{eff}_{9}
\bar{s}\gamma_\mu (1-\gamma_5) b \, \bar{\ell} \gamma^\mu \ell +
C_{10}  \bar{s} \gamma_\mu (1-\gamma_5) b \, \bar{\ell}
\gamma^\mu
\gamma_5 \ell \nnb \\
&-&  2 m_b C^{eff}_{7} {1\over q^2} \bar{s} i \sigma_{\mu\nu} q^{\nu}
(1+\gamma_5) b \, \bar{\ell} \gamma^\mu \ell \Bigg]~, \eea
where $G_{F}$ is the Fermi coupling constant, $\alpha_{em}$ is the fine structure constant at $Z$ mass scale, $V_{ij}$ are elements of the Cabibbo-Kobayashi-Maskawa (CKM) matrix; and 
the $C^{eff}_{7}$, $C^{eff}_{9}$ and $C_{10}$ are the Wilson coefficients. Considering the contributions of the  new  operators coming from  the new interactions 
 induced by the NHBs exchanged diagrams, the supersymmetric effective Hamiltonian   can be written as 
\bea \label{HeffSUSY}\label{heff-susy} {\cal H}^{eff}_{SUSY} &=& {G_F \alpha_{em} V_{tb}
V_{ts}^\ast \over 2\sqrt{2} \pi} \Bigg[ C^{eff}_{9}
\bar{s}\gamma_\mu (1-\gamma_5) b \, \bar{\ell} \gamma^\mu \ell + C^{\prime_eff}_{9} \bar{s} \gamma_\mu (1+\gamma_5) b \, \bar{\ell} \gamma^\mu \ell \nnb \\
&+& C_{10}  \bar{s} \gamma_\mu (1-\gamma_5) b \, \bar{\ell} \gamma^\mu \gamma_5 \ell + C^{\prime}_{10} \bar{s} \gamma_\mu (1+\gamma_5) b \, \bar{\ell} \gamma^\mu \gamma_5 \ell \nnb\\ 
&-& 2 m_b C^{eff}_{7} {1\over q^2} \bar{s} i \sigma_{\mu\nu} q^{\nu} (1+\gamma_5) b \, \bar{\ell} \gamma^\mu \ell -  2 m_b C^{\prime_eff}_{7} {1\over q^2} \bar{s} i \sigma_{\mu\nu} q^{\nu} (1-\gamma_5) b \, \bar{\ell} \gamma^\mu \ell \nnb \\
&+& C_{Q_1} \bar{s}  (1+\gamma_5) b \, \bar{\ell} \ell + C^{\prime}_{Q_1} \bar{s} (1-\gamma_5) b \, \bar{\ell} \ell  \nnb \\
&+& C_{Q_2} \bar{s} (1+\gamma_5) b \, \bar{\ell} \gamma_5 \ell + C^{\prime}_{Q_2} \bar{s} (1-\gamma_5) b \, \bar{\ell} \gamma_5 \ell \Bigg]~, \eea
where  the new Wilson coefficients, $C_{Q_1}$ and $C_{Q_2}$  exist in the all considered SUSY models, while the  primed coefficients  only appear in SUSY SO(10) scenario.


The amplitude is obtained by sandwiching the new effective Hamiltonian between the initial and final baryonic states, i.e., 
\bea\label{amplitude}
{\cal M}_{SUSY}^{ \Lambda_b \rightarrow \Lambda \ell^+ \ell^-} = \langle \Lambda(p_{\Lambda}) \mid{\cal H}^{eff}_{SUSY}\mid \Lambda_b(p_{\Lambda_b})\rangle,
\eea
where $p_{\Lambda_b}$ and $p_{\Lambda}$ are momenta of the $\Lambda_b$ and $\Lambda$ baryons, respectively. To proceed, we need to calculate the following matrix elements  parametrized in terms of twelve form factors  in full theory:
\bea\label{SMtransmatrix} \langle
\Lambda(p_{\Lambda}) \mid  \bar s \gamma_\mu (1-\gamma_5) b \mid \Lambda_b(p_{\Lambda_b})\rangle\es
\bar {u}_\Lambda(p_{\Lambda}) \Bigg[\gamma_{\mu}f_{1}(q^{2})+{i}
\sigma_{\mu\nu}q^{\nu}f_{2}(q^{2}) + q^{\mu}f_{3}(q^{2}) \nnb \\
\ek \gamma_{\mu}\gamma_5
g_{1}(q^{2})-{i}\sigma_{\mu\nu}\gamma_5q^{\nu}g_{2}(q^{2})
- q^{\mu}\gamma_5 g_{3}(q^{2})
\vphantom{\int_0^{x_2}}\Bigg] u_{\Lambda_{b}}(p_{\Lambda_b})~,\nnb \\
\eea
\bea\label{Susytransmatrix} \langle
\Lambda(p_{\Lambda}) \mid  \bar s \gamma_\mu (1+\gamma_5) b \mid \Lambda_b(p_{\Lambda_b})\rangle\es
\bar {u}_\Lambda(p_{\Lambda}) \Bigg[\gamma_{\mu}f_{1}(q^{2})+{i}
\sigma_{\mu\nu}q^{\nu}f_{2}(q^{2}) + q^{\mu}f_{3}(q^{2}) \nnb \\
&+& \gamma_{\mu}\gamma_5
g_{1}(q^{2})+{i}\sigma_{\mu\nu}\gamma_5q^{\nu}g_{2}(q^{2})
+ q^{\mu}\gamma_5 g_{3}(q^{2})
\vphantom{\int_0^{x_2}}\Bigg] u_{\Lambda_{b}}(p_{\Lambda_b})~,\nnb \\
\eea
\bea\label{SMtransmatrix2}
\langle \Lambda(p_{\Lambda})\mid \bar s i \sigma_{\mu\nu}q^{\nu} (1+ \gamma_5)
b \mid \Lambda_b(p_{\Lambda_b})\rangle \es\bar{u}_\Lambda(p_{\Lambda})
\Bigg[\gamma_{\mu}f_{1}^{T}(q^{2})+{i}\sigma_{\mu\nu}q^{\nu}f_{2}^{T}(q^{2})+
q^{\mu}f_{3}^{T}(q^{2}) \nnb \\
\ar \gamma_{\mu}\gamma_5
g_{1}^{T}(q^{2})+{i}\sigma_{\mu\nu}\gamma_5q^{\nu}g_{2}^{T}(q^{2})
+ q^{\mu}\gamma_5 g_{3}^{T}(q^{2})
\vphantom{\int_0^{x_2}}\Bigg] u_{\Lambda_{b}}(p_{\Lambda_b})~,\nnb \\
\eea
\bea\label{Susytransmatrix2}
\langle \Lambda(p_{\Lambda})\mid \bar s i \sigma_{\mu\nu}q^{\nu} (1- \gamma_5)
b \mid \Lambda_b(p_{\Lambda_b})\rangle \es\bar{u}_\Lambda(p_{\Lambda})
\Bigg[\gamma_{\mu}f_{1}^{T}(q^{2})+{i}\sigma_{\mu\nu}q^{\nu}f_{2}^{T}(q^{2})+
q^{\mu}f_{3}^{T}(q^{2}) \nnb \\
\ek \gamma_{\mu}\gamma_5
g_{1}^{T}(q^{2})-{i}\sigma_{\mu\nu}\gamma_5q^{\nu}g_{2}^{T}(q^{2})
- q^{\mu}\gamma_5 g_{3}^{T}(q^{2})
\vphantom{\int_0^{x_2}}\Bigg] u_{\Lambda_{b}}(p_{\Lambda_b})~,\nnb \\
\eea
\bea\label{SUSYtransmatrix3}\langle
\Lambda(p_{\Lambda}) \mid  \bar s (1+\gamma_5) b \mid \Lambda_b(p_{\Lambda_b})\rangle\es
{1 \over m_b} \bar {u}_\Lambda(p_{\Lambda}) \Bigg[{\not\!q}f_{1}(q^{2})+{i}
q^{\mu} \sigma_{\mu\nu}q^{\nu}f_{2}(q^{2}) + q^{2}f_{3}(q^{2}) \nnb \\
\ek {\not\!q}\gamma_5
g_{1}(q^{2})-{i}q^{\mu}\sigma_{\mu\nu}\gamma_5q^{\nu}g_{2}(q^{2})
- q^{2}\gamma_5 g_{3}(q^{2})
\vphantom{\int_0^{x_2}}\Bigg] u_{\Lambda_{b}}(p_{\Lambda_b})~,\nnb \\
\eea
and,
\bea\label{SUSYtransmatrix4}\langle
\Lambda(p_{\Lambda}) \mid  \bar s (1-\gamma_5) b \mid \Lambda_b(p_{\Lambda_b})\rangle\es
{1 \over m_b} \bar {u}_\Lambda(p_{\Lambda}) \Bigg[{\not\!q} f_{1}(q^{2})+{i}
q^{\mu} \sigma_{\mu\nu}q^{\nu}f_{2}(q^{2}) + q^{2}f_{3}(q^{2}) \nnb \\
\ar {\not\!q}\gamma_5
g_{1}(q^{2})+{i}q^{\mu} \sigma_{\mu\nu}\gamma_5q^{\nu}g_{2}(q^{2})
+ q^{2}\gamma_5 g_{3}(q^{2})
\vphantom{\int_0^{x_2}}\Bigg] u_{\Lambda_{b}}(p_{\Lambda_b})~,\nnb \\
\eea
where  $q^2$ is the transformed momentum squared;  and the $u_{\Lambda_b}$ and $u_{\Lambda}$ are spinors of the initial and final baryons. 
In the meantime, the $f^{(T)}_i$ and $g^{(T)}_i$ with $i=1,2$ and $3$ are transition form factors. 

Using the above transition matrix elements in terms of form factors, we find the supersymmetric amplitude as
\bea\label{amplitude1}
&&{\cal M}_{SUSY}^{ \Lambda_b \rightarrow \Lambda \ell^+ \ell^-} = {G_F \alpha_{em} V_{tb}V_{ts}^\ast \over 2\sqrt{2} \pi} \Bigg\{\nnb \\
&&\Big[{\bar u}_\Lambda ({p}_{\Lambda}) ( \gamma_{\mu}[{\cal A}_1 R + {\cal B}_1 L]+ {i}\sigma_{\mu\nu} q^{\nu}[{\cal A}_2 R + {\cal B}_2 L] + q^{\mu} [{\cal A}_3 R + {\cal B}_3 L]) u_{\Lambda_{b}}(p_{\Lambda_b}) \Big] \, (\bar{\ell} \gamma^\mu \ell)\nnb \\
&+& \Big[{\bar u}_\Lambda ({p}_{\Lambda})( \gamma_{\mu}[{\cal D}_1 R + {\cal E}_1 L]+ {i}\sigma_{\mu\nu} q^{\nu}[{\cal D}_2 R +{\cal E}_2 L] + q^{\mu} [{\cal D}_3 R + {\cal E}_3 L]) u_{\Lambda_{b}}(p_{\Lambda_b}) \Big] \,(\bar{\ell} \gamma^\mu \gamma_5 \ell) \nnb \\
&+& \Big[{\bar u}_\Lambda ({p}_{\Lambda})( {\not\!q} [{\cal G}_1 R + {\cal H}_1 L]+ {i} q^{\mu} \sigma_{\mu\nu} q^{\nu}[{\cal G}_2 R + {\cal H}_2 L] + q^2 [{\cal G}_3 R + {\cal H}_3 L]) u_{\Lambda_{b}}(p_{\Lambda_b}) \Big] \, (\bar{\ell} \ell)\nnb \\
&+& \Big[{\bar u}_\Lambda ({p}_{\Lambda})( {\not\!q}[{\cal K}_1 R + {\cal S}_1 L]+ {i} q^{\mu} \sigma_{\mu\nu}q^{\nu}[{\cal K}_2 R + {\cal S}_2 L] + q^2 [{\cal K}_3 R + {\cal S}_3 L]) u_{\Lambda_{b}}(p_{\Lambda_b}) \Big] \, (\bar{\ell} \gamma_5 \ell) \Bigg\} ~,\nnb \\
\eea
where $R=(1+\gamma_5)/2$ and $L=(1-\gamma_5)/2$ and the calligraphic coefficients  are found as

\bea \label{coef-decay-rate} {\cal A}_{1} \es f_1 C_{9}^{eff+} - g_1 C_{9}^{eff-} - 2 m_b {1\over q^2} \Big[f_1^T C_{7}^{eff+} + g_1^T C_{7}^{eff-} \Big] ,~ {\cal A}_{2} = {\cal A}_1 ( 1 \rar 2 ),~ {\cal A}_{3} = {\cal A}_1 \ga 1 \rar 3 \dr ~,\nnb 
\eea
\bea
{\cal B}_{1} \es f_1 C_{9}^{eff+} + g_1 C_{9}^{eff-} - 2 m_b {1\over q^2} \Big[ f_1^T C_{7}^{eff+} - g_1^T C_{7}^{eff-} \Big],~ {\cal B}_{2} = {\cal B}_1 \ga 1 \rar 2 \dr, ~{\cal B}_{3} = {\cal B}_1 \ga 1 \rar 3 \dr ~,\nnb 
\eea
\bea
{\cal D}_{1} = f_1 C_{10}^{+} - g_1 C_{10}^{-},~~~~~~~~~~~~~~~~ {\cal D}_{2} =  {\cal D}_1 \ga 1 \rar 2 \dr,~~~~~~~~~ {\cal D}_{3} = {\cal D}_1 \ga 1 \rar 3 \dr,~~~~~~~~~~~~~~~~~~~~~~~~~~~\nnb 
\eea
\bea
{\cal E}_{1} = f_1 C_{10}^{+} + g_1 C_{10}^{-},~~~~~~~~~~~~~~~~~ {\cal E}_{2} = {\cal E}_1 \ga 1 \rar 2 \dr,~~~~~~~~~~ {\cal E}_{3} = {\cal E}_1 \ga 1 \rar 3 \dr,~~~~~~~~~~~~~~~~~~~~~~~~~~~\nnb  
\eea
\bea
{\cal G}_{1} = {1\over m_b}  \Big[ f_1 C_{Q_1}^{+} - g_1 C_{Q_1}^{-} \Big],~~~~~~~~~ {\cal G}_{2} = {\cal G}_1 \ga 1 \rar 2 \dr,~~~~~~~~~ {\cal G}_{3} = {\cal G}_1 \ga 1 \rar 3 \dr,~~~~~~~~~~~~~~~~~~~~~~~~~~~\nnb 
\eea
\bea
{\cal H}_{1} = {1\over m_b} \Big[ f_1 C_{Q_1}^{+} + g_1 C_{Q_1}^{-}  \Big],~~~~~~~~ {\cal H}_{2} = {\cal H}_1 \ga 1 \rar 2 \dr,~~~~~~~~~ {\cal H}_{3} =  {\cal H}_1 \ga 1 \rar 3 \dr,~~~~~~~~~~~~~~~~~~~~~~~~~~\nnb 
\eea
\bea
{\cal K}_{1} = {1\over m_b} \Big[ f_1 C_{Q_2}^{+} - g_1 C_{Q_2}^{-} \Big],~~~~~~~~~ {\cal K}_{2} =  {\cal K}_1 \ga 1 \rar 2 \dr,~~~~~~~~~ {\cal K}_{3} =  {\cal K}_1 \ga 1 \rar 3 \dr,~~~~~~~~~~~~~~~~~~~~~~~~~~~\nnb 
\eea
\bea
{\cal S}_{1} = {1\over m_b} \Big[ f_1 C_{Q_2}^{+} + g_1 C_{Q_2}^{-} \Big],~~~~~~~~~~ {\cal S}_{2} = {\cal S}_1 \ga 1 \rar 2 \dr,~~~~~~~~~~ {\cal S}_{3} = {\cal S}_1 \ga 1 \rar 3 \dr,~~~~~~~~~~~~~~~~~~~~~~~~~~~\nnb\\
\eea
with
\bea
C_{9}^{eff+} \es C_{9}^{eff}+C^{\prime_eff}_{9},~~~~~~~~~~~~~~~~~~~~~~~~~~~~~~~~~~~~ C_{9}^{eff-} = C_{9}^{eff}-C^{\prime_eff}_{9}~, \nnb
\eea
\bea
C_{7}^{eff+} \es C_{7}^{eff}+C^{\prime_eff}_{7},~~~~~~~~~~~~~~~~~~~~~~~~~~~~~~~~~~~~ C_{7}^{eff-} = C_{7}^{eff}-C^{\prime_eff}_{7}~, \nnb
\eea
\bea
C_{10}^{+}~~~ \es C_{10}~~+~~C^{\prime}_{10},~~~~~~~~~~~~~~~~~~~~~~~~~~~~~~~~~~~~C_{10}^{-}~~~ = C_{10}~~-~~C^{\prime}_{10}~, \nnb
\eea
\bea
C_{Q_1}^{+}~~~ \es C_{Q_1}~~+~~C^{\prime}_{Q_1},~~~~~~~~~~~~~~~~~~~~~~~~~~~~~~~~~~~~C_{Q_1}^{-}~~ = C_{Q_1}~~-~~C^{\prime}_{Q_1}~, \nnb
\eea
\bea
C_{Q_2}^{+}~~~ \es C_{Q_2}~~+~~C^{\prime}_{Q_2},~~~~~~~~~~~~~~~~~~~~~~~~~~~~~~~~~~~~C_{Q_2}^{-}~~ = C_{Q_2}~~-~~C^{\prime}_{Q_2}~. \nnb\\
\eea


\section{Differential decay rate, branching fraction and FBA}

\subsection{The differential decay rate }

In this part, we calculate the differential decay rate for the decay channel under consideration. Using the aforementioned amplitude, we find the supersymmetric differential decay rate in terms of form factors in
full theory as:
\bea\label{DDR} \frac{d^2\Gamma}{d\hat
sdz}(z,\hat s) = \frac{G_F^2\alpha^2_{em} m_{\Lambda_b}}{16384
\pi^5}| V_{tb}V_{ts}^*|^2 v \sqrt{\lambda(1,r,\hat s)} \, \Bigg[{\cal
T}_{0}(\hat s)+{\cal T}_{1}(\hat s) z +{\cal T}_{2}(\hat s)
z^2\Bigg]~, \nnb\\ \label{dif-decay}
\eea
where $\lambda=\lambda(1,r,\hat s)=(1-r-\hat s)^2-4r\hat s$ is the usual triangle function with $\hat s= q^2/m^2_{\Lambda_b}$, $r= m^2_{\Lambda}/m^2_{\Lambda_b}$ and $v=\sqrt{1-\frac{4 m_\ell^2}{q^2}}$.
 Here also $z=\cos\theta$ with $\theta$ is the angle between momenta of the lepton $l^+$ and the  $\Lambda_b$ in the center of mass of leptons.
The calligraphic, ${\cal T}_{0}(\hat s)$, ${\cal T}_{1}(\hat s)$ and ${\cal T}_{2}(\hat s)$ functions are obtained  as:
 \bea {\cal T}_{0}(\hat s) \es 32 m_\ell^2
m_{\Lambda_b}^4 \hat s (1+r-\hat s) \Big( \vel {\cal D}_{3} \ver^2 +
\vel {\cal E}_{3} \ver^2 \Big) \nnb \\
\ar 64 m_\ell^2 m_{\Lambda_b}^3 (1-r-\hat s) \, \mbox{\rm Re} \Big[{\cal D}_{1}^\ast
{\cal E}_{3} + {\cal D}_{3}
{\cal E}_1^\ast \Big] \nnb \\
\ar 64 m_{\Lambda_b}^2 \sqrt{r} (6 m_\ell^2 - m_{\Lambda_b}^2 \hat s)
{\rm Re} \Big[{\cal D}_{1}^\ast {\cal E}_{1}\Big] \nnb\\ 
\ar 64 m_\ell^2 m_{\Lambda_b}^3 \sqrt{r} \Bigg\{ 2 m_{\Lambda_b} \hat s
{\rm Re} \Big[{\cal D}_{3}^\ast {\cal E}_{3}\Big] + (1 - r + \hat s)
{\rm Re} \Big[{\cal D}_{1}^\ast {\cal D}_{3} + {\cal E}_{1}^\ast {\cal E}_{3}\Big]\Bigg\} \nnb \\
\ar 32 m_{\Lambda_b}^2 (2 m_\ell^2 + m_{\Lambda_b}^2 \hat s) \Bigg\{ (1
- r + \hat s) m_{\Lambda_b} \sqrt{r} \,
\mbox{\rm Re} \Big[{\cal A}_{1}^\ast {\cal A}_{2} + {\cal B}_{1}^\ast {\cal B}_{2}\Big] \nnb \\
\ek m_{\Lambda_b} (1 - r - \hat s) \, \mbox{\rm Re} \Big[{\cal A}_{1}^\ast {\cal B}_{2} +
{\cal A}_{2}^\ast {\cal B}_{1}\Big] - 2 \sqrt{r} \Big( \mbox{\rm Re} \Big[{\cal A}_{1}^\ast {\cal B}_{1}\Big] +
m_{\Lambda_b}^2 \hat s \,
\mbox{\rm Re} \Big[{\cal A}_{2}^\ast {\cal B}_{2}\Big] \Big) \Bigg\} \nnb 
\eea
\bea
\ar 8 m_{\Lambda_b}^2 \Bigg\{ 4 m_\ell^2 (1 + r - \hat s) +
m_{\Lambda_b}^2 \Big[(1-r)^2 - \hat s^2 \Big]
\Bigg\} \Big( \vel {\cal A}_{1} \ver^2 +  \vel {\cal B}_{1} \ver^2 \Big) \nnb \\
\ar 8 m_{\Lambda_b}^4 \Bigg\{ 4 m_\ell^2 \Big[ \lambda + (1 + r -
\hat s) \hat s \Big] + m_{\Lambda_b}^2 \hat s \Big[(1-r)^2 - \hat s^2 \Big]
\Bigg\} \Big( \vel {\cal A}_{2} \ver^2 +  \vel {\cal B}_{2} \ver^2 \Big) \nnb \\
\ek 8 m_{\Lambda_b}^2 \Bigg\{ 4 m_\ell^2 (1 + r - \hat s) -
m_{\Lambda_b}^2 \Big[(1-r)^2 - \hat s^2 \Big]
\Bigg\} \Big( \vel {\cal D}_{1} \ver^2 +  \vel {\cal E}_{1} \ver^2 \Big) \nnb\\
\ar 8 m_{\Lambda_b}^5 \hat s v^2 \Bigg\{ - 8 m_{\Lambda_b} \hat s \sqrt{r}\,
\mbox{\rm Re} \Big[{\cal D}_{2}^\ast {\cal E}_{2}\Big] +
4 (1 - r + \hat s) \sqrt{r} \, \mbox{\rm Re}\Big[{\cal D}_{1}^\ast {\cal D}_{2}+{\cal E}_{1}^\ast {\cal E}_{2}\Big]\nnb \\
\ek 4 (1 - r - \hat s) \, \mbox{\rm Re}\Big[{\cal D}_{1}^\ast {\cal E}_{2}+{\cal D}_{2}^\ast {\cal E}_{1}\Big] +
m_{\Lambda_b} \Big[(1-r)^2 -\hat s^2\Big] \Big( \vel {\cal D}_{2} \ver^2 + \vel
{\cal E}_{2} \ver^2 \Big) \Bigg\} \nnb \\
\ek 8 m_{\Lambda_b}^4 \Bigg\{ 4 m_\ell \Big[(1-r)^2 -\hat s(1+r) \Big]\, \mbox{\rm Re} \Big[{\cal D}_{1}^\ast {\cal K}_{1}+{\cal E}_{1}^\ast {\cal S}_{1}\Big] \nnb \\ 
\ar (4 m_\ell^2 - m_{\Lambda_b}^2 \hat s) \Big[(1-r)^2 -\hat s(1+r) \Big]\, \Big( \vel {\cal G}_{1} \ver^2 + \vel {\cal H}_{1} \ver^2 \Big) \nnb \\
\ar 4 m_{\Lambda_b}^2 \sqrt{r} \hat s^2 (4 m_\ell^2 - m_{\Lambda_b}^2 \hat s) \, \mbox{\rm Re}\Big[{\cal G}_{3}^\ast {\cal H}_{3}\Big] \Bigg\} \nnb \\
\ek 8 m_{\Lambda_b}^5 \hat s \Bigg\{2 \sqrt{r} (4 m_\ell^2 - m_{\Lambda_b}^2 \hat s) \,(1 - r + \hat s) \, \mbox{\rm Re}\Big[{\cal G}_{1}^\ast {\cal G}_{3}+{\cal H}_{1}^\ast {\cal H}_{3}\Big] \nnb \\
\ar 4 m_\ell \sqrt{r}(1 - r + \hat s) \mbox{\rm Re}\Big[{\cal D}_{1}^\ast {\cal K}_{3}+{\cal E}_{1}^\ast {\cal S}_{3}+{\cal D}_{3}^\ast {\cal K}_{1}+{\cal E}_{3}^\ast {\cal S}_{1}\Big] \nnb \\
\ar 4 m_\ell (1 - r - \hat s) \mbox{\rm Re}\Big[{\cal D}_{1}^\ast {\cal S}_{3}+{\cal E}_{1}^\ast {\cal K}_{3}+{\cal D}_{3}^\ast {\cal S}_{1}+{\cal E}_{3}^\ast {\cal K}_{1}\Big] \nnb \\
\ar 2 (1 - r - \hat s) (4 m_\ell^2 - m_{\Lambda_b}^2 \hat s) \, \mbox{\rm Re}\Big[{\cal G}_{1}^\ast {\cal H}_{3}+{\cal H}_{1}^\ast {\cal G}_{3}\Big] \nnb \\
\ek m_{\Lambda_b} \Big[(1-r)^2 -\hat s(1+r) \Big] \Big( \vel {\cal K}_{1} \ver^2 +  \vel {\cal S}_{1} \ver^2 \Big) \Bigg\} \nnb \\
\ek 32 m_{\Lambda_b}^4 \sqrt{r} \hat s \Bigg\{ 2 m_\ell \mbox{\rm Re}\Big[{\cal D}_{1}^\ast {\cal S}_{1}+{\cal E}_{1}^\ast {\cal K}_{1}\Big]+(4 m_\ell^2 - m_{\Lambda_b}^2 \hat s) \,\mbox{\rm Re}\Big[{\cal G}_{1}^\ast {\cal H}_{1}\Big] \Bigg\} \nnb \\
\ar 8 m_{\Lambda_b}^6 \hat s^2 \Bigg\{ 4 \sqrt{r} \,\mbox{\rm Re}\Big[{\cal K}_{1}^\ast {\cal S}_{1}\Big]+2 m_{\Lambda_b} \sqrt{r} (1 - r + \hat s) \mbox{\rm Re}\Big[{\cal K}_{1}^\ast {\cal K}_{3}+{\cal S}_{1}^\ast {\cal S}_{3}\Big] \nnb \\
&+& 2 m_{\Lambda_b} (1 - r - \hat s) \mbox{\rm Re}\Big[{\cal K}_{1}^\ast {\cal S}_{3}+{\cal S}_{1}^\ast {\cal K}_{3}\Big] \nnb \\
\ek (4 m_\ell^2 - m_{\Lambda_b}^2 \hat s) (1 + r - \hat s) \Big( \vel {\cal G}_{3} \ver^2 +  \vel {\cal H}_{3} \ver^2 \Big) \nnb \\
\ek 4 m_\ell (1 + r - \hat s) \mbox{\rm Re}\Big[{\cal D}_{3}^\ast {\cal K}_{3}+{\cal E}_{3}^\ast {\cal S}_{3}\Big]- 8 m_\ell \sqrt{r} \mbox{\rm Re}\Big[{\cal D}_{3}^\ast {\cal S}_{3}+{\cal E}_{3}^\ast {\cal K}_{3}\Big]\Bigg\} \nnb \\
\ar 8 m_{\Lambda_b}^8 \hat s^3 \Bigg\{ (1 + r - \hat s) \Big( \vel {\cal K}_{3} \ver^2 +  \vel {\cal S}_{3} \ver^2 \Big) + 4 \sqrt{r} \mbox{\rm Re}\Big[{\cal K}_{3}^\ast {\cal S}_{3}\Big]\Bigg\}, \nnb \\
\eea
\bea {\cal T}_{1}(\hat s) &=& -32
m_{\lb}^4 m_\ell \sqrt{\lambda} v (1 - r)
Re\Big({\cal A}_{1}^* {\cal G}_{1}+{\cal B}_{1}^* {\cal H}_{1}\Big)\nn\\
&-&16 m_{\lb}^4\s1 v \sqrt{\lambda} 
\Bigg\{ 2 Re\Big({\cal A}_{1}^* {\cal D}_{1}\Big)-2Re\Big({\cal B}_{1}^* {\cal E}_{1}\Big)\nn\\
&+&2 m_{\lb} Re\Big({\cal B}_{1}^* {\cal D}_{2}-{\cal B}_{2}^* {\cal D}_{1}+{\cal A}_{2}^* {\cal E}_{1}-{\cal A}_{1}^*{\cal E}_{2}\Big)\nn\\
&+&2 m_{\lb} m_\ell Re\Big({\cal A}_{1}^* {\cal H}_{3}+{\cal B}_{1}^* {\cal G}_{3}-{\cal A}_{2}^* {\cal H}_{1}-{\cal B}_{2}^*{\cal G}_{1}\Big)\Bigg\}\nn\\
&+&32 m_{\lb}^5 \s1~ v \sqrt{\lambda} \Bigg\{
m_{\lb} (1-r)Re\Big({\cal A}_{2}^* {\cal D}_{2} -{\cal B}_{2}^* {\cal E}_{2}\Big)\nn\\
&+& \sqrt{r} Re\Big({\cal A}_{2}^* {\cal D}_{1}+{\cal A}_{1}^* {\cal D}_{2}-{\cal B}_{2}^*{\cal E}_{1}-{\cal B}_{1}^* {\cal E}_{2}\Big)\nn\\
&-& \sqrt{r} m_\ell Re\Big({\cal A}_{1}^* {\cal G}_{3}+{\cal B}_{1}^* {\cal H}_{3}+{\cal A}_{2}^*{\cal G}_{1}+{\cal B}_{2}^* {\cal H}_{1}\Big)\Bigg\} \nn\\
&+&32 m_{\lb}^6 m_\ell \sqrt{\lambda} v \hat s^2 Re\Big({\cal A}_{2}^* {\cal G}_{3}+{\cal B}_{2}^* {\cal H}_{3}\Big),\nn\\
\eea\
\bea {\cal T}_{2}(\hat s) \es - 8 m_{\Lambda_b}^4 v^2 \lambda \Big(\vel {\cal A}_{1} \ver^2 + \vel {\cal B}_{1} \ver^2 + \vel {\cal D}_{1} \ver^2 + \vel {\cal E}_{1} \ver^2 \Big) \nnb \\
\ar 8 m_{\Lambda_b}^6 \hat s v^2 \lambda \Big( \vel {\cal A}_{2} \ver^2 + \vel
{\cal B}_{2} \ver^2 + \vel {\cal D}_{2} \ver^2 + \vel {\cal E}_{2} \ver^2 \Big) ~.\nn\\ 
\eea\

In order to obtain the differential decay rate only in terms of $ \hat s$, we fulfill integrate  Eq.\eqref{DDR} over $z$  in the interval $[-1,1]$. As a result, we get
 \bea
\frac{d\Gamma}{d \hat s} (\hat s)= \frac{G_F^2\alpha^2_{em} m_{\Lambda_b}}{8192
\pi^5}| V_{tb}V_{ts}^*|^2 v \sqrt{\lambda} \, \Bigg[{{\cal T}_0(\hat s)
+\frac{1}{3} {\cal T}_2(\hat s)}\Bigg]~. \label{decayrate} \eea

\subsection{The differential branching ratio }
Using the differential decay rate, in this subsection, we numerically analyze the differential branching ratio and calculate the values of the branching ratios at different lepton channels.
For this aim, we need sum inputs which we would like to present them here. In  Table 1, we present the masses  \cite{PDG} as well as the lifetime of the initial baryon  \cite{PDG},
 some constants and elements of the CKM matrix.
\begin{table}[ht]
\centering
\rowcolors{1}{lightgray}{white}
\begin{tabular}{cc}
\hline \hline
   Some Input Parameters  &  Values    
           \\
\hline \hline
$ m_e $              &   $ 0.00051  $ $GeV$ \\
$ m_\mu $            &   $ 0.1056   $ $GeV$ \\
$ m_\tau $           &   $ 1.776 $  $GeV$ \\
$ m_b $              &   $ 4.8 $ $GeV$  \\
$ m_{\Lambda_b} $    &   $ 5.620 $ $GeV$   \\
$ m_{\Lambda} $      &   $ 1.1156 $ $GeV$   \\
$ \tau_{\Lambda_b} $ &  $ 1.425\times 10^{-12} $ $s$  \\
$ \hbar  $           &   $ 6.582\times 10^{-25} GeV s $   \\
$ G_{F} $            &   $ 1.17\times 10^{-5} $ $GeV^{-2}$ \\
$ \alpha_{em} $      &   $ 1/137 $   \\
$ | V_{tb}V_{ts}^*| $ &   $ 0.041 $   \\
 \hline \hline
 
\end{tabular}
\caption{The values of some input parameters ​​used in the analysis.}
\end{table}

The main inputs in our calculations are form factors. These form factors are calculated via light cone QCD sum rules in full theory in \cite{form-factors}.
The fit function for the form factors  $f_{1}$, $f_{2}$, $f_{3}$, $g_{1}$, $g_{2}$, $g_{3}$, $f^T_{2}$, $f^T_{3}$, $g^T_{2}$ and $g^T_{3}$ is given   as \cite{form-factors}:
\bea\label{formfactors}
 f^{(T)}_{i}(q^2)[g^{(T)}_{i}(q^2)]=\frac{a}{\Bigg(1-\ds\frac{q^2}{m_{fit}^2}\Bigg)}+
\frac{b}{\Bigg(1-\ds\frac{q^2}{m_{fit}^2}\Bigg)^2}~,
\eea
 where the fit parameters
$a,~b$ and $m_{fit}^2$ as well as the values of the related form factors
 at $q^2=0$ in full theory are given in Table 2.
\begin{table}[h!]
\centering
\rowcolors{1}{lightgray}{white}
\begin{tabular}{ccccc}
\hline \hline
                & $\mbox{a} $ & $ \mbox{b} $ & $m_{fit}^2$ & \mbox{form factors at} $q^2=0$  \\
\hline \hline
 $ f_1    $        & $ -0.046 $ & $   0.368  $  & $ 39.10 $ & $ 0.322 \pm 0.112$ \\
 $ f_2     $       & $  0.0046 $ & $ -0.017 $  & $  26.37 $ & $-0.011 \pm 0.004 $  \\
 $ f_3      $      & $  0.006 $ & $  -0.021 $  & $  22.99 $ & $-0.015 \pm 0.005 $ \\
 $ g_1       $     & $ -0.220 $ & $   0.538 $  & $  48.70 $ & $0.318 \pm 0.110 $\\
 $ g_2       $     & $  0.005  $ & $ -0.018 $  & $  26.93 $ & $-0.013 \pm 0.004$   \\
 $ g_3        $    & $  0.035  $ & $ -0.050 $ &   $ 24.26 $ & $-0.014 \pm 0.005$ \\
 $ f_2^{T}    $    & $ -0.131 $ & $   0.426 $   & $ 45.70 $ & $0.295 \pm 0.105 $ \\
 $ f_3^{T}    $    & $ -0.046  $ &  $ 0.102 $   & $ 28.31 $ & $0.056 \pm 0.018 $ \\
 $ g_2^{T}     $   & $ -0.369 $  & $  0.664 $   & $ 59.37 $  & $0.294 \pm 0.105$ \\
 $ g_3^{T}     $   & $ -0.026 $ & $  -0.075 $   & $ 23.73 $ & $-0.101 \pm 0.035$ \\
\hline \hline
\end{tabular}
\caption{The parameters in the fit function of the form factors  $f_{1}$, $f_{2}$, $f_{3}$, $g_{1}$, $g_{2}$, $g_{3}$, $f^T_{2}$, $f^T_{3}$, $g^T_{2}$ and $g^T_{3}$ as well as their values
 at $q^2=0$  in full theory \cite{form-factors}.}
\end{table}
\begin{table}[ht]
\centering
\rowcolors{1}{lightgray}{white}
\begin{tabular}{ccccc}
\hline \hline
           & $\mbox{c} $ & $ m_{fit}^{' 2}$ & $ m_{fit}^{'' 2} $ & \mbox{form factors at} $q^2=0$ \\
          \hline \hline
$ f_1^{T} $ &   $ -1.191 $  & $ 23.81 $   & $ 59.96  $ &  $0     \pm 0.0$      \\
$ g_1^{T} $ &   $ -0.653 $ & $ 24.15 $  & $ 48.52   $  & $0     \pm 0.0$ \\
 \hline \hline
 \end{tabular}
\caption{The parameters in the fit function of the form factors $f^T_{1}$ and  $g^T_{1}$ as well as their values
 at $q^2=0$ in full theory \cite{form-factors}.}
\end{table}
Furthermore, the fit function of the form factors $f^T_{1}$ and $g^T_{1}$ is given by \cite{form-factors}: 
\bea\label{frmfctft1gt1}
 f^T_{1}(q^2)[g^T_{1}(q^2)]=\frac{c}{\Bigg(1-\ds\frac{q^2}{m_{fit}^{' 2}}\Bigg)}-
\frac{c}{\Bigg(1-\ds\frac{q^2}{m_{fit}^{'' 2}}\Bigg)^2}~, \eea
 where, the parameters $c$, $m^{'2}_{fit}$ and $m^{''2}_{fit}$ as well as the values of the corresponding form factors
 at $q^2=0$ are  presented in Table 3.

In our numerical analysis, it is important to emphasize that the Wilson coefficient $C^{eff}_{9}$ has been taken to contain also   the long distance (LD) effects coming from the charmonium resonances. 
These effects are parameterized using the  Breit-Weigner ansatz as \cite{Buchalla,Beneke,Khodjamirian}:
\begin{eqnarray}
Y_{LD}&=&\frac{3\pi }{\alpha^{2}}C^{(0)}  \sum_{i=1}^6 \kappa_i \, \frac{\Gamma
(V_{i}\rightarrow \ell^+ \ell^-)m_{V_{i}}}{m_{V_{i}}^{2}-q^{2}-im_{V_{i}}\Gamma _{V_{i}}},
\label{LD}
\end{eqnarray}%
where, $ C^{(0)}=0.362 $ and $\kappa_i$ are the phenomenological factors. Here, $m_{V_{i}}$ and $\Gamma _{V_{i}}$ are the masses and decay rates of the vector  charmonia, respectively. In the present work,
 we only take into account the two lowest resonances that are $J/\psi(1s) $ and $\psi(2s)$. The phenomenological factors have also been chosen as $\kappa_1\cong 1$ and $\kappa_2\cong 2$. The masses, 
branching fractions and total decay widths related to the considered resonances are ​​given in  Table 4. 
\begin{table}[ht]
\centering
\rowcolors{1}{lightgray}{white}
\begin{tabular}{cccc}
\hline \hline
   Family of $J/\psi $   &  $Mass [GeV]$ &   $\Gamma(V_{i}\rightarrow \ell^+ \ell^-)$ & $\Gamma _{V_{i}}$ 
           \\
\hline \hline
$ J/\psi(1s)  $  &   $ 3.096  $  &  $5.55\times 10^{-6}$ & $92.9\times 10^{-6}$  \\
$ \psi(2s)$  &   $ 3.686  $  &  $2.35\times 10^{-6}$ & $304\times 10^{-6}$  \\
 \hline \hline
 \end{tabular}
\caption{The values of masses, branching fractions and total decay widths related to the resonances $J/\psi(1s) $ and $\psi(2s)$  \cite{PDG}.}
\end{table}

Considering the above mentioned resonances from  $J/\psi$ family, we divide the  allowed physical regions into the following three regions in the case of  the electron and  muon as final leptons: 
\begin{eqnarray*}
Region ~I   &;&4m_{l}^{2}\leq q^{2}\leq (m_{J/\psi(1s)}-0.02)^{2},\\
Region ~II &;&(m_{J/\psi(1s) }+0.02)^{2}\leq q^{2}\leq(m_{\psi(2s)}-0.02)^{2},\\
Region ~III &;&(m_{\psi(2s)}+0.02)^{2}\leq q^{2}\leq(m_{\varLambda_{b}}-m_{\varLambda})^{2}.
\end{eqnarray*}%

In the case of  $\tau$, we have the following two regions:
\begin{eqnarray*}
Region ~I &;&4m_{\tau }^{2}\leq q^{2}\leq (m_{\psi(2s)}-0.02)^{2},\\
Region ~II &;&(m_{\psi(2s)}+0.02)^{2}\leq q^{2}\leq(m_{\varLambda_{b}}-m_{\varLambda})^{2}.\\
\end{eqnarray*}%

Finally, we would like to present the numerical values of the Wilson coefficients used in numerical calculations in Table 5.
\begin{table}[ht]
\centering
\rowcolors{1}{lightgray}{white}
\begin{tabular}{cccccc}
\hline \hline
\mbox{{\small Coefficient}} &\mbox{{\small SM}}&\mbox{{\small SUSY I}}&\mbox{{\small SUSY II}}&\mbox{{\small SUSY III}}&\mbox{{\small SUSY SO(10)($A_{0}=-1000$)}}
              \\
\hline
 ${\small C^{eff}_{7}} $   & ${\small -0.313} $&  ${\small +0.376}$ & $ {\small +0.376 }  $   & $ {\small -0.376} $    &$ -0.219 $                  \\
 ${\small C^{eff}_{9}}$    & $ {\small 4.334} $&   ${\small 4.767}$ &  $ {\small 4.767 }   $  & $  {\small 4.767}  $   &  ${\small 4.275}   $                \\
 ${\small C_{10} }    $    & $ {\small -4.669}$ &  ${\small -3.735}$ & $ {\small -3.735}   $   & $ {\small -3.735} $    & ${\small -4.732}  $                 \\
 ${\small C_{Q_1} }    $    & $ {\small 0}     $&   ${\small 0}     $& $ {\small 6.5 (16.5)} $ & $ {\small 1.2 (4.5)}$  & $ {\small 0.106+0i(1.775+0.002i)}$  \\
 ${\small C_{Q_2}  }   $    & $ {\small 0}     $&   ${\small 0}    $& ${\small -6.5 (-16.5)}$ & ${\small -1.2 (-4.5)} $& ${\small -0.107+0i(-1.797-0.002i)}$ \\
 ${\small C^{\prime_eff}_{7}}  $   & $ {\small 0} $ & $ {\small 0} $ & $ {\small 0} $ & $ {\small 0} $ & $ {\small 0.039+0.038i}$ \\
$ {\small C^{\prime_eff}_{9}}  $   & $ {\small 0} $ & $ {\small 0} $ & $ {\small 0}$ & $ {\small 0} $ & $ {\small 0.011+0.072i }$\\
 ${\small C^{\prime}_{10} }    $   & $ {\small 0} $ & $ {\small 0} $ & $ {\small 0} $ & $ {\small 0} $ & ${\small -0.075-0.67i} $ \\
 ${\small C^{\prime}_{Q_1} }    $   & $ {\small 0} $ & $ {\small 0} $ & $ {\small 0} $ & $ {\small 0} $ & ${\small -0.247+0.242i(-4.148+4.074i)}$ \\
 ${\small C^{\prime}_{Q_2}   }  $   & $ {\small 0} $ & $ {\small 0} $ & $ {\small 0} $ & $ {\small 0} $ & ${\small -0.25+0.246i(-4.202+4.128i)} $ \\
\hline\hline
\end{tabular}
\caption{The Wilson coefficients used in numerical calculations \cite{Aslam2,SUSY,Wil.coef,Aslam3}. In the values containing parentheses, the values inside the parentheses  stand for the $\tau$ lepton, while the 
values outside belong to the $e$ and $\mu$ cases. The other values (without parentheses) refer to all leptons.}
\end{table}
\begin{figure}[h!]
\centering
\begin{tabular}{ccc}
\epsfig{file=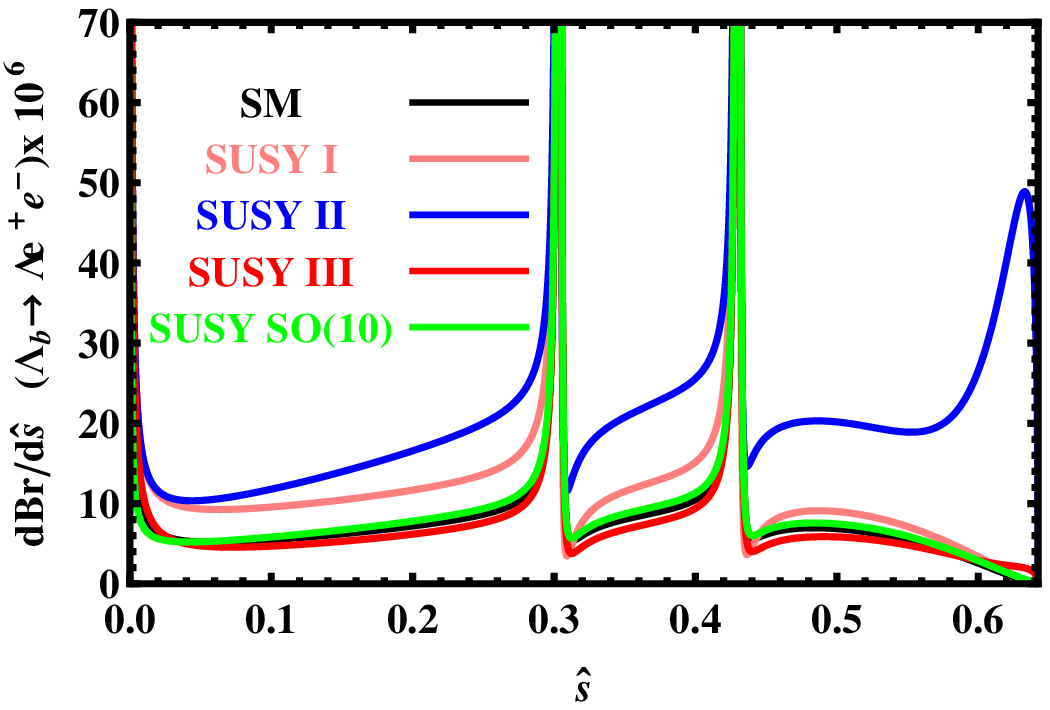,width=0.45\linewidth,clip=} &
\epsfig{file=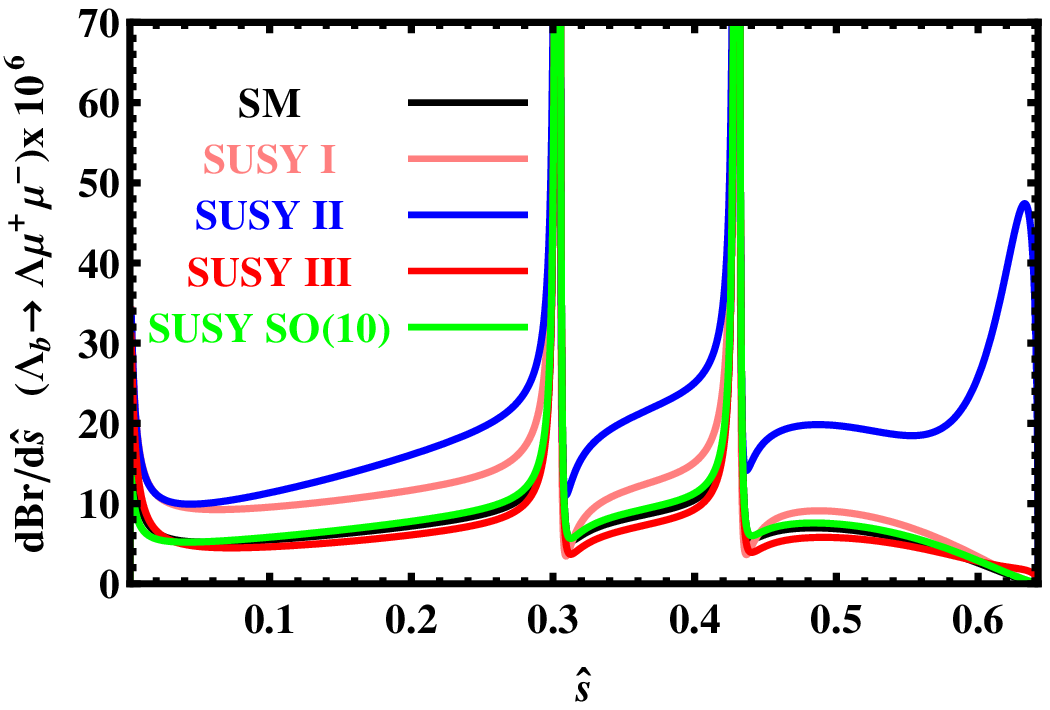,width=0.45\linewidth,clip=} 
\end{tabular}
\caption{The dependence of the differential branching ratio on  $\hat s$  for the $\Lambda_{b}\rightarrow \Lambda e^{+} e^{-}$ and $\Lambda_{b}\rightarrow \Lambda \mu^{+} \mu^{-}$ transitions in  SM and different
 SUSY models using the central values of form factors.}
\end{figure}
\begin{figure}[h!]
\centering
\begin{tabular}{ccc}
\epsfig{file=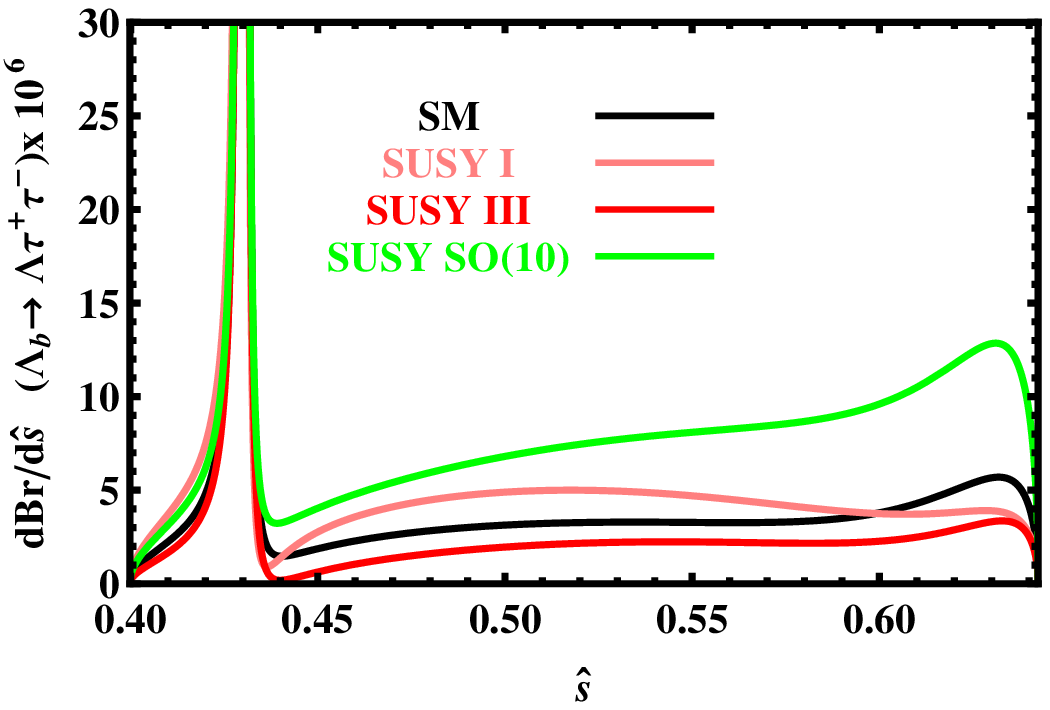,width=0.45\linewidth,clip=} &
\epsfig{file=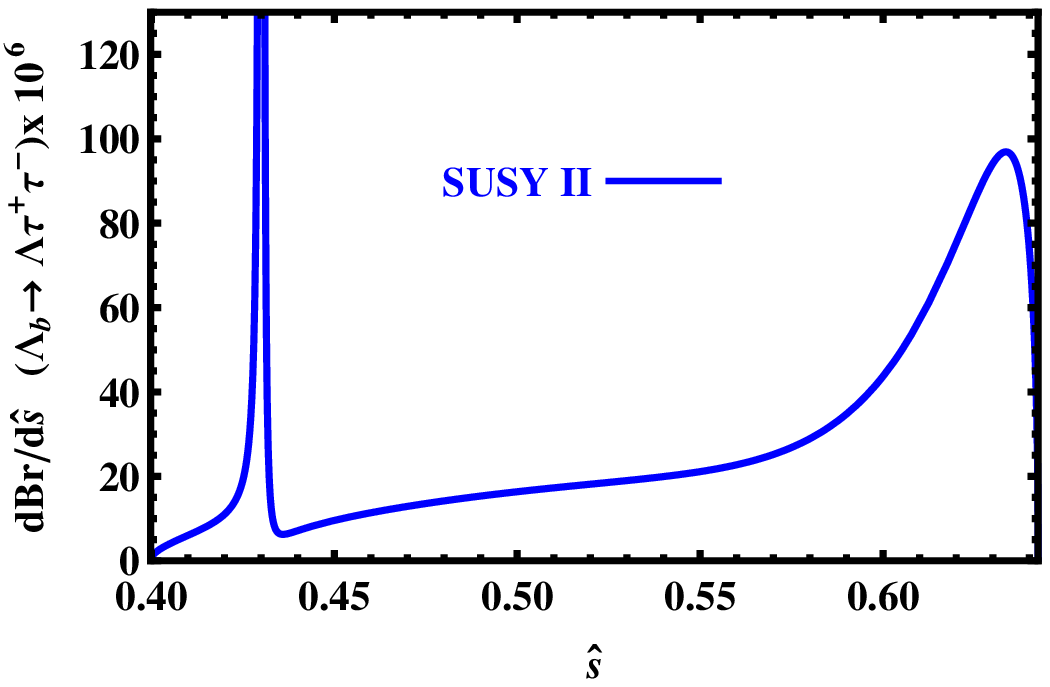,width=0.45\linewidth,clip=} 
\end{tabular}
\caption{The dependence of the differential branching ratio on  $\hat s$  for the $\Lambda_{b}\rightarrow \Lambda \tau^{+} \tau^{-}$ transition in  SM and different SUSY models using the central values of form factors.}
\end{figure}

Having given all the  inputs,  we now  present the dependence of the differential branching ratio on  $\hat s$ for the $e, ~\mu $ and $\tau $ leptons in the SM and different SUSY scenarios in Figures 1 and 2.
 From these figures which are plotted considering the central values of the form factors, we see that
\begin{itemize}
\item in all lepton channels, the predictions of the SUSY II deviate maximally from those of the SM and other considered SUSY models. In the case of $\tau$, this deviation reaches to approximately one order of magnitude.
\item As far as the $e$ and $\mu$ are concerned, the results obtained via the SUSY I have also considerable deviation from the predictions of the other models.
\item In the case of $\tau$ as final lepton, we see sizable differences between all models' predictions. The nearest results to the SM correspond to the SUSY I and III.
\end{itemize}
\begin{table}[ht]
\centering
\rowcolors{1}{lightgray}{white}
\begin{tabular}{cccc}
\hline \hline
 \mbox{\textit{BR}($\Lambda_{b}\rightarrow \Lambda e^{+} e^{-}$)} &\mbox{\textit{ Region I}}&\mbox{\textit{ Region II}}&\mbox{\textit{ Region III}}
                \\
\hline \hline
 SM                         & $2.86\times 10^{-6}$ & $1.12\times 10^{-6}$ &  $0.81\times 10^{-6}$  \\

 SUSY I                      & $4.57\times 10^{-6}$ &  $1.56\times 10^{-6}$ &  $0.99\times 10^{-6}$   \\

 SUSY II                     & $5.66\times 10^{-6}$ &  $2.69\times 10^{-6}$ & $2.61\times 10^{-6}$    \\

 SUSY III                    & $2.93\times 10^{-6}$ &  $0.98\times 10^{-6}$ &  $0.66\times 10^{-6}$  \\

 SUSY SO(10)                & $2.65\times 10^{-6}$ &  $1.20\times 10^{-6}$ &  $1.09\times 10^{-6}$   \\
 
\hline \hline
\end{tabular}
\caption{The central values  of branching ratio for $\Lambda_{b}\rightarrow \Lambda e^{+} e^{-}$ decay channel at different regions in SM and different SUSY models.}
\end{table}
\begin{table}[ht]
\centering
\rowcolors{1}{lightgray}{white}
\begin{tabular}{cccc}
\hline \hline
\mbox{\textit{BR}($\Lambda_{b}\rightarrow \Lambda \mu^{+} \mu^{-}$)}& \mbox{\textit{ Region I}} & \mbox{\textit{ Region II}}& \mbox{\textit{ Region III}}
                  \\
\hline \hline
 SM                         & $2.25\times 10^{-6}$ & $1.12\times 10^{-6}$ &  $0.81\times 10^{-6}$  \\

 SUSY I                      & $3.69\times 10^{-6}$ &  $1.56\times 10^{-6}$ &  $0.99\times 10^{-6}$   \\

 SUSY II                     & $4.65\times 10^{-6}$ &  $2.63\times 10^{-6}$ & $2.55\times 10^{-6}$    \\

 SUSY III                    & $2.04\times 10^{-6}$ &  $0.97\times 10^{-6}$ &  $0.65\times 10^{-6}$  \\

 SUSY SO(10)                & $2.33\times 10^{-6}$ &  $1.20\times 10^{-6}$ &  $1.09\times 10^{-6}$  \\
\hline\hline
\end{tabular}
\caption{The central values of branching ratio  for $\Lambda_{b}\rightarrow \Lambda \mu^{+} \mu^{-}$ decay channel at different regions in SM and different SUSY models.}
\end{table}
\begin{table}[ht]
\centering
\rowcolors{1}{lightgray}{white}
\begin{tabular}{ccc}
\hline \hline
 \mbox{\textit{BR}($\Lambda_{b}\rightarrow \Lambda \tau^{+} \tau^{-}$)} &\mbox{\textit{ Region I}}&\mbox{\textit{Region II }}
                \\
\hline \hline
 SM                         & $0.87\times 10^{-7}$ & $3.84\times 10^{-7}$   \\

 SUSY I                      & $1.35\times 10^{-7}$ &  $5.55\times 10^{-7}$  \\

 SUSY II                     & $2.01\times 10^{-7}$ &  $2.44\times 10^{-6}$  \\

 SUSY III                    & $0.72\times 10^{-7}$ &  $2.13\times 10^{-7}$  \\

 SUSY SO(10)                & $1.12\times 10^{-7}$ &  $1.61\times 10^{-6}$  \\
 
\hline \hline
\end{tabular}
\caption{The central values of branching ratio  for $\Lambda_{b}\rightarrow \Lambda \tau^{+} \tau^{-}$ decay channel at different regions in  SM and different SUSY models.}
\end{table}
\subsection{The  branching ratio }

Integrating the differential branching ratio over $\hat s$ in the considered regions and taking into account the central values of the form factors, we find the branching ratios for various models as presented in Tables 6, 7 and 8 for different lepton channels.
 A quick glance at these Tables leads to the following results:                                                                                                                                                                                                     
  \begin{itemize}
 \item as it is expected the  values of the branching ratio decrease when going from the $e$ to $\tau$.
\item The order of branching ratios indicates that these channels are accessible at the LHC. Note that as we have already mentioned this decay channel has been observed by CDF Collaboration at Fermilab in $\mu$
channel \cite{CDF-lambdab}.
\item All SUSY models have predictions considerably different than those of the SM in all regions and at all lepton channels.
\item The maximum deviation from the SM results belongs to the SUSY II model. When considering the numerical values, the maximum deviation of  the SUSY II result from the SM prediction corresponds to the  region II for 
$\tau$ channel. In this case, the result of the SUSY II is approximately $6$ times greater than that of the SM.
%
\end{itemize}


\subsection{The FBA }

The lepton forward-backward asymmetry (${\cal A}_{FB}$) is defined as:
\bea {\cal A}_{FB} = \frac{N_f-N_b}{N_f+N_b}.
 \eea
where $N_f$ is the number of moving particles to forward direction and $N_b$ is the number of moving particles to backward direction. 
In technique language, the lepton FBA is written in terms of the differential decay rate as:
\bea {\cal A}_{FB} (\hat s)=
\frac{\ds{\int_0^1\frac{d^{2}\Gamma}{d\hat{s}dz}}(z,\hat s)\,dz -
\ds{\int_{-1}^0\frac{d^{2}\Gamma}{d\hat{s}dz}}(z,\hat s)\,dz}
{\ds{\int_0^1\frac{d^{2}\Gamma}{d\hat{s}dz}}(z,\hat s)\,dz +
\ds{\int_{-1}^0\frac{d^{2}\Gamma}{d\hat{s}dz}}(z,\hat s)\,dz}~. 
\eea

\begin{figure}[h!]
\centering
\begin{tabular}{ccc}
\epsfig{file=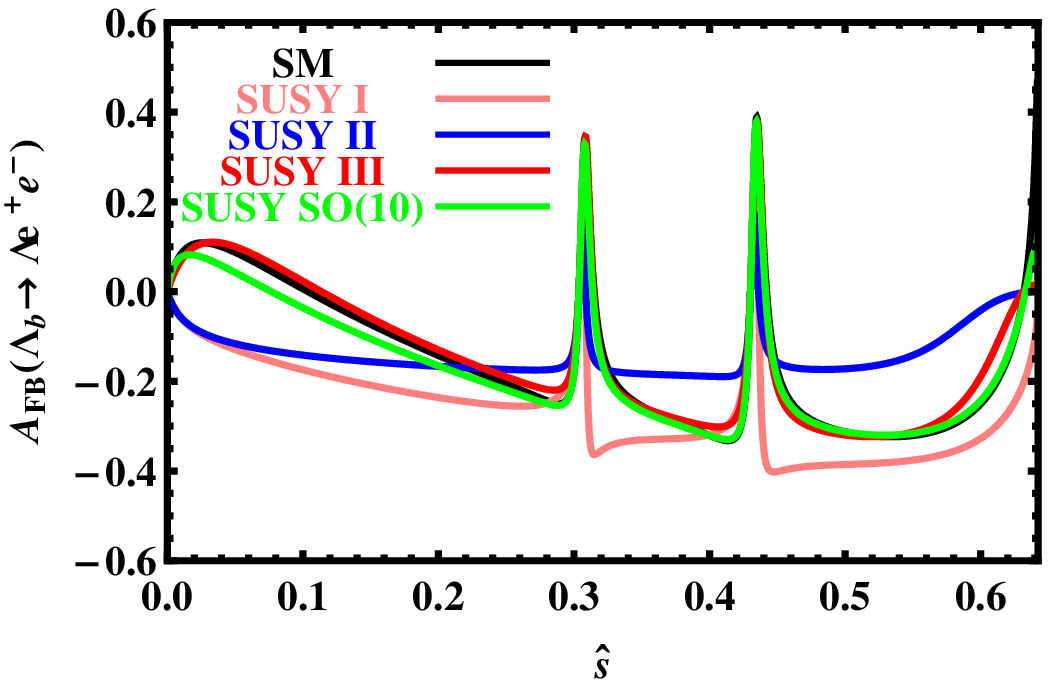,width=0.45\linewidth,clip=} &
\epsfig{file=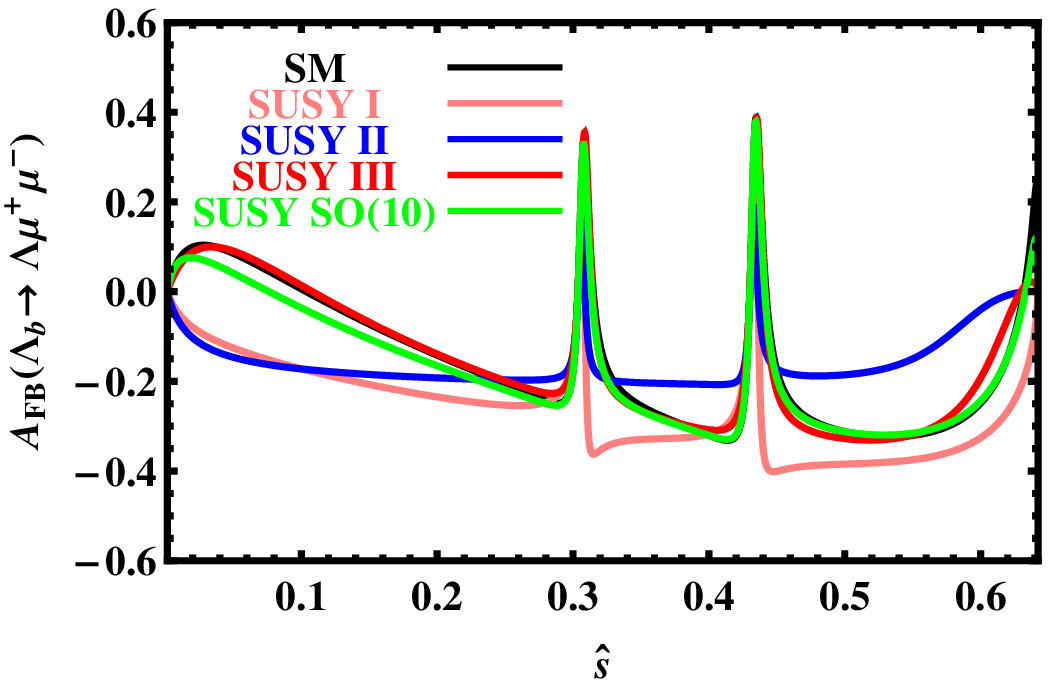,width=0.45\linewidth,clip=} 
\end{tabular}
\caption{The dependence of the FBA on  $\hat s$  for  $\Lambda_{b}\rightarrow \Lambda e^{+} e^{-}$ and $\Lambda_{b}\rightarrow \Lambda \mu^{+} \mu^{-}$ transitions in  SM and different SUSY scenarios using the central values of form factors.}
\end{figure}

\begin{figure}[h!]
\centering
\begin{tabular}{ccc}
\epsfig{file=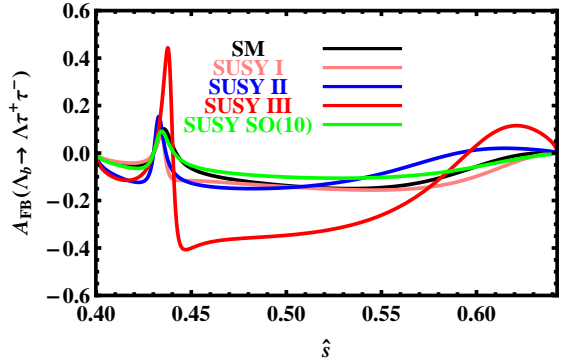,width=0.45\linewidth,clip=}
\end{tabular}
\caption{The dependence of the FBA on  $\hat s$  for $\Lambda_{b}\rightarrow \Lambda \tau^{+} \tau^{-}$ transition in SM and different SUSY scenarios using the central values of form factors.}
\end{figure}
Using this definition,  we  plot the dependence of the lepton FBA on  $\hat s$  for  $e, ~\mu $ and $\tau $ channels  in  the SM and different SUSY models in Figures 3 and 4.
 From these figures which are also plotted considering the central values of the form factors, it is clear that,
\begin{itemize}
\item in the case of the $e$ and $\mu$ channels, the SUSY I and II behave different than the other models. In these channels for the small values of the  $\hat s$, the maximum deviation belongs to the SUSY I, however,
for higher values of the  $\hat s$ the maximum deviation corresponds to the SUSY II.
\item In $\tau $ channel, the maximum deviation from the SM prediction belongs to the SUSY III.
\item The zero points of the FBA  in different SUSY models move slightly  to the  left compare to the SM predictions. In some regions, the SUSY I, II and III have different signs with the SM predictions.
\item The SUSY SO(10) represents overall the closest results to the SM predictions.
\end{itemize}

\subsection{The physical quantities under consideration taking into account the uncertainties of the form factors}
 \begin{figure}[h!]
\centering
\begin{tabular}{ccc}
\epsfig{file=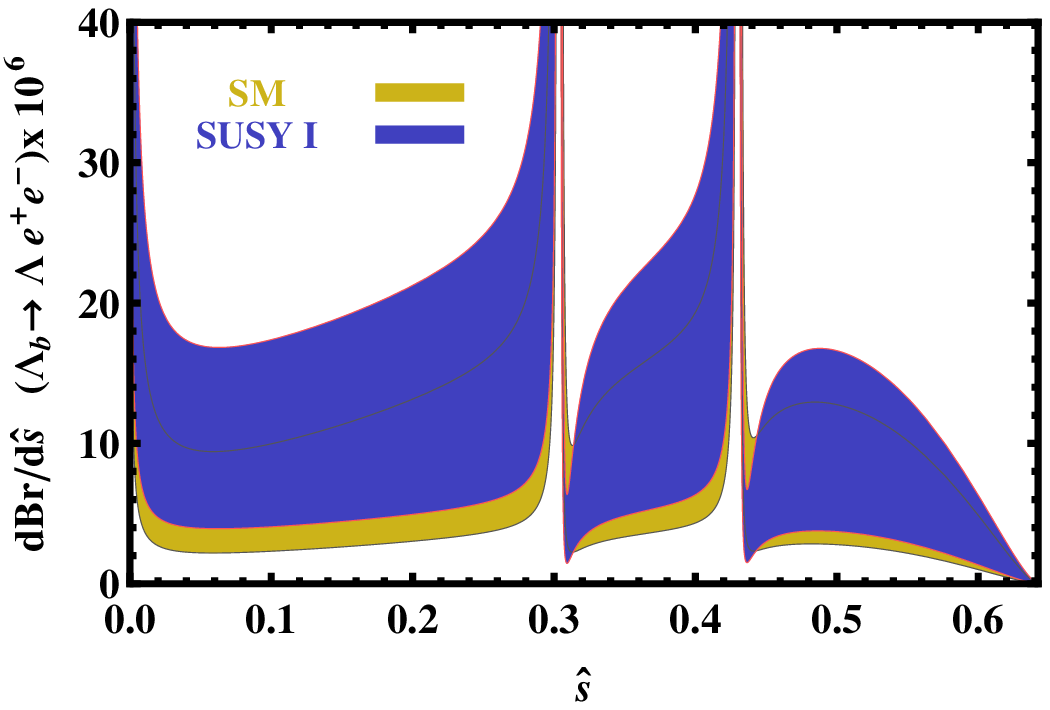,width=0.35\linewidth,clip=} &
\epsfig{file=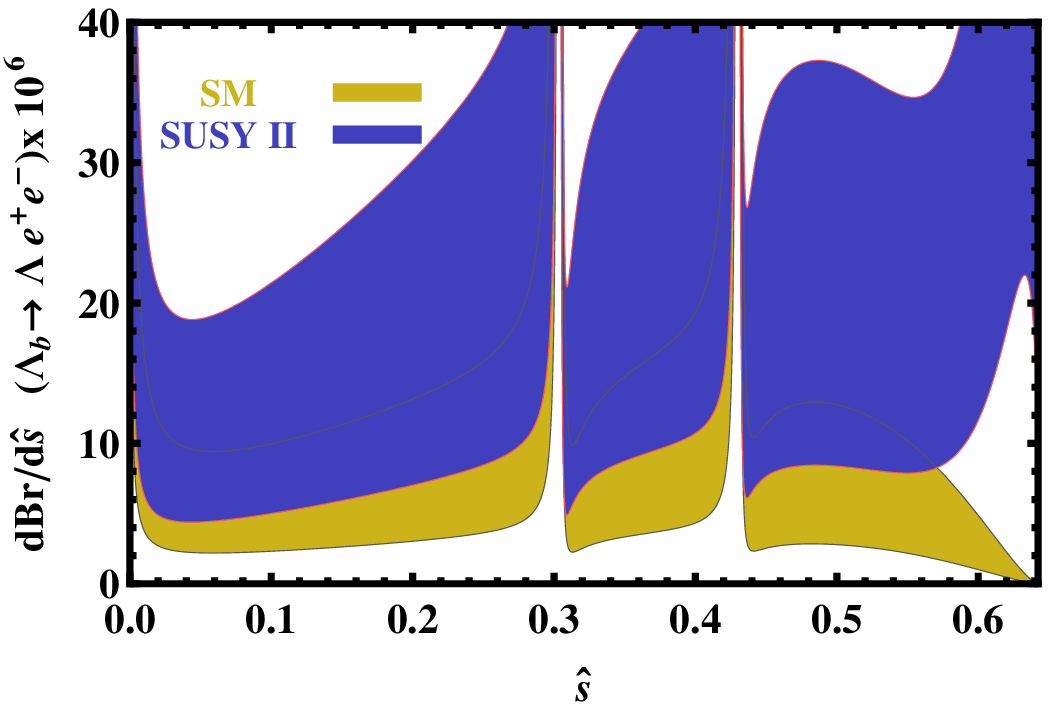,width=0.35\linewidth,clip=}
\end{tabular}
\begin{tabular}{ccc}
\epsfig{file=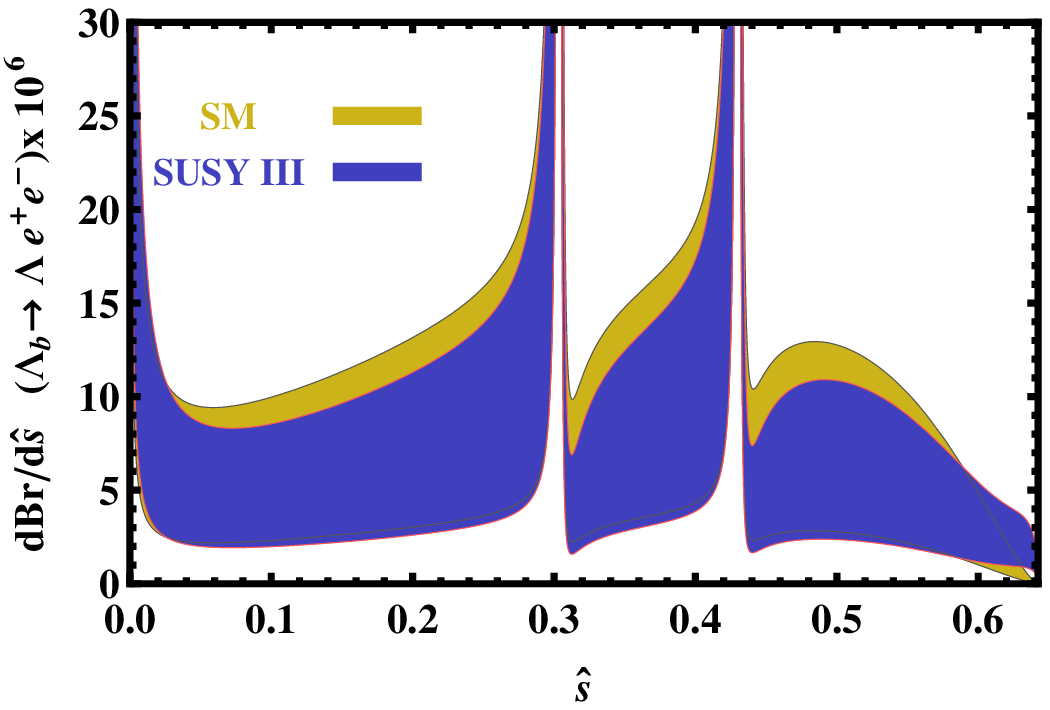,width=0.35\linewidth,clip=} &
\epsfig{file=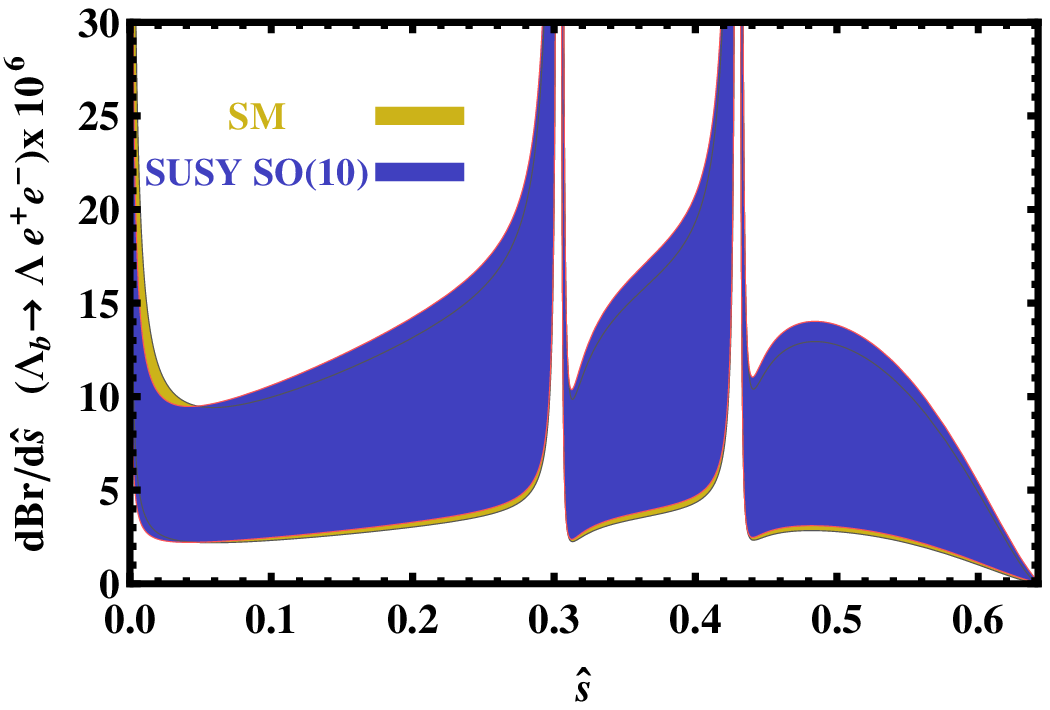,width=0.35\linewidth,clip=}
\end{tabular}
\caption{Comparison of the results of differential branching ratio  with respect to  $\hat s$  for the $\Lambda_{b}\rightarrow \Lambda e^{+} e^{-}$  transition obtained  from different
 SUSY models with that of the   SM  considering the errors of form factors.}
\end{figure}
\begin{figure}[h!]
\centering
\begin{tabular}{ccc}
\epsfig{file=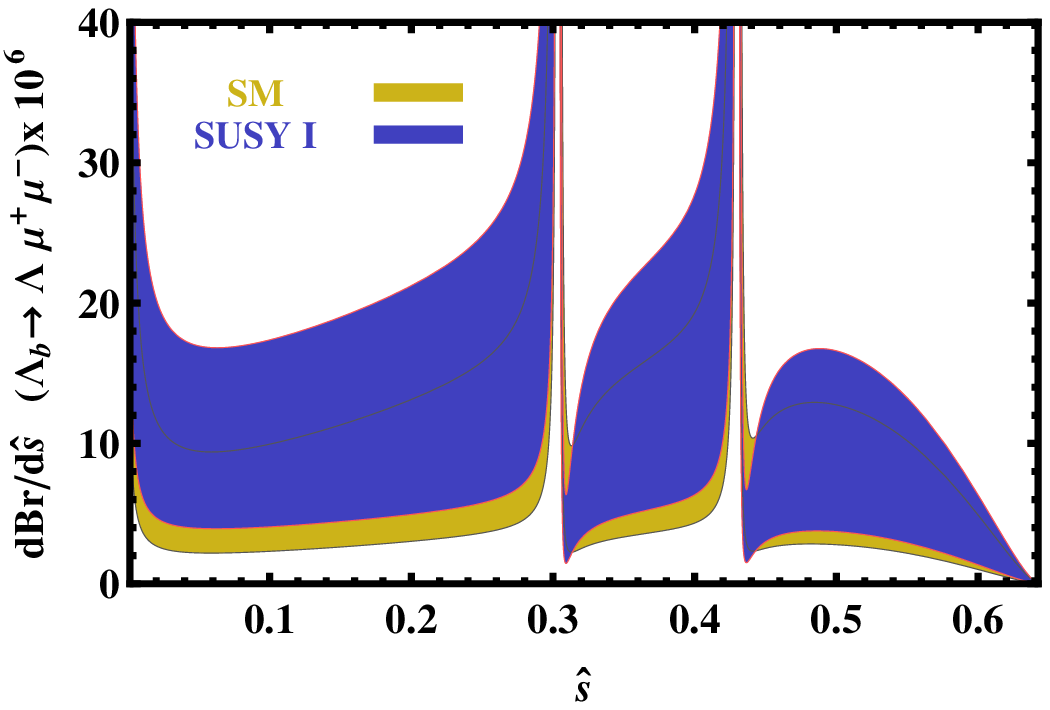,width=0.35\linewidth,clip=} &
\epsfig{file=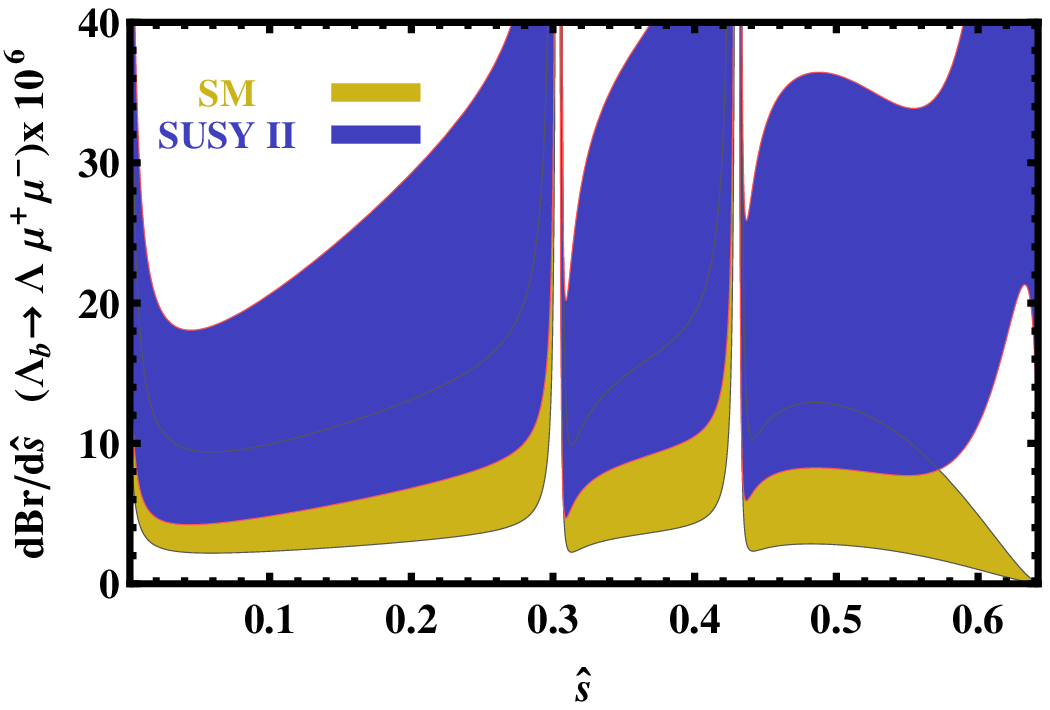,width=0.35\linewidth,clip=}
\end{tabular}
\begin{tabular}{ccc}
\epsfig{file=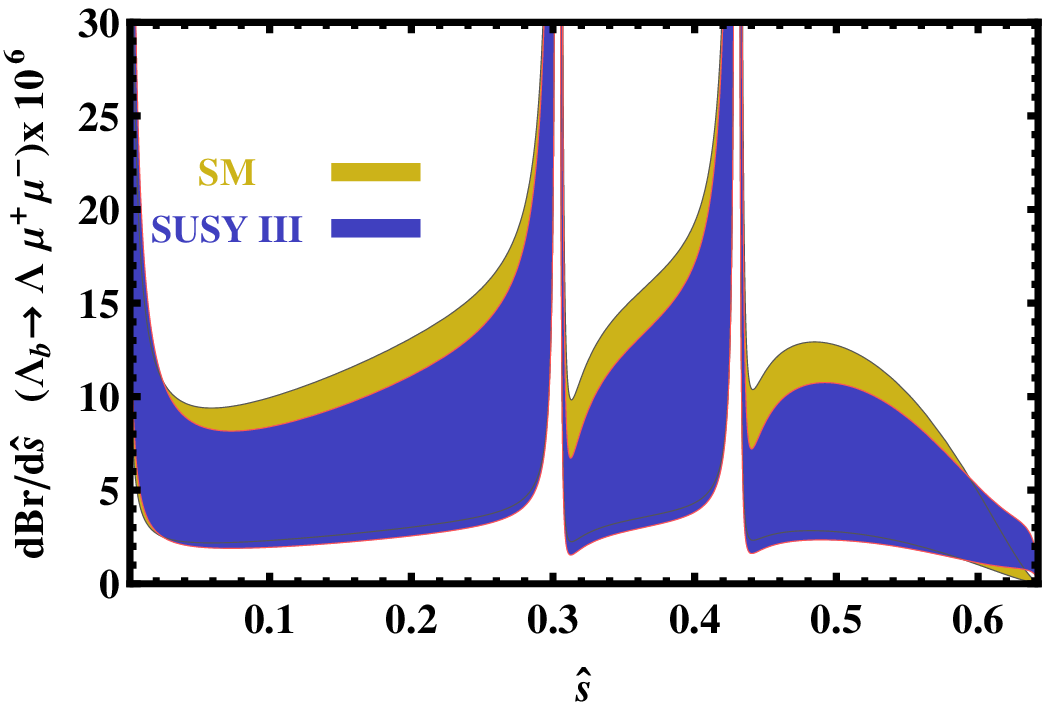,width=0.35\linewidth,clip=} &
\epsfig{file=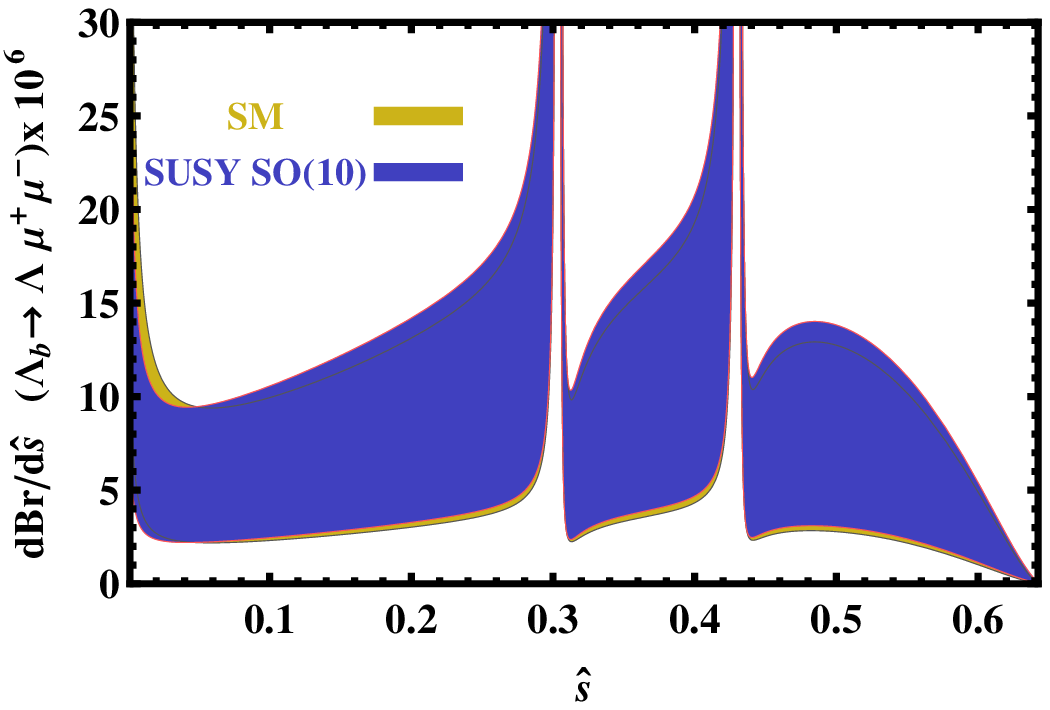,width=0.35\linewidth,clip=}
\end{tabular}
\caption{The same as Figure 5 but for $\mu$.}
\end{figure}
\begin{figure}[h!]
\centering
\begin{tabular}{ccc}
\epsfig{file=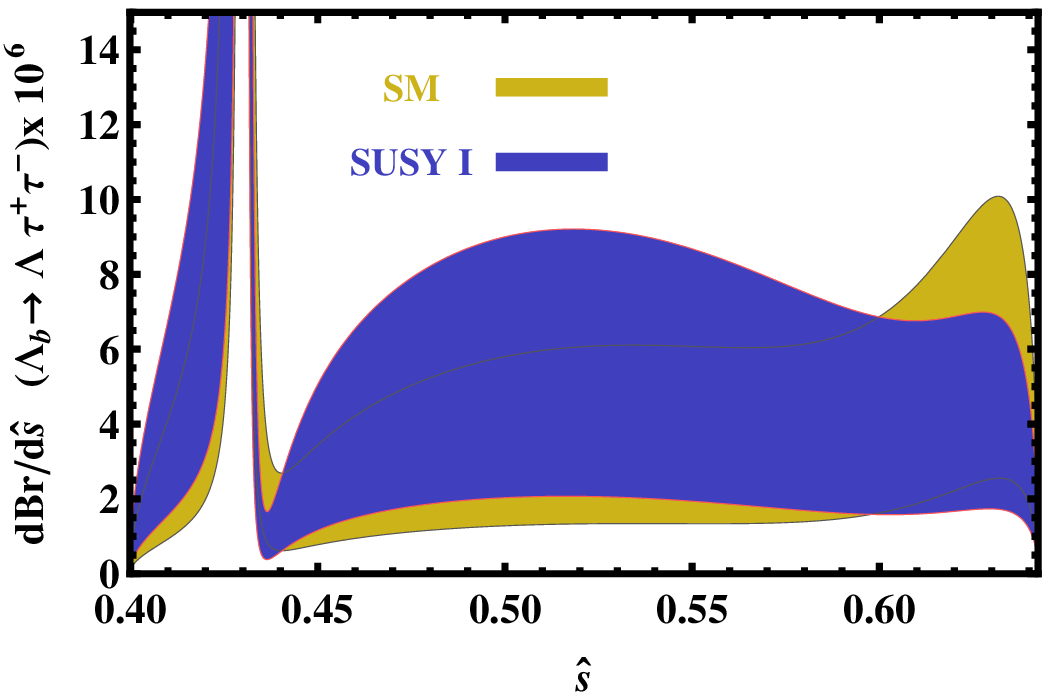,width=0.35\linewidth,clip=} &
\epsfig{file=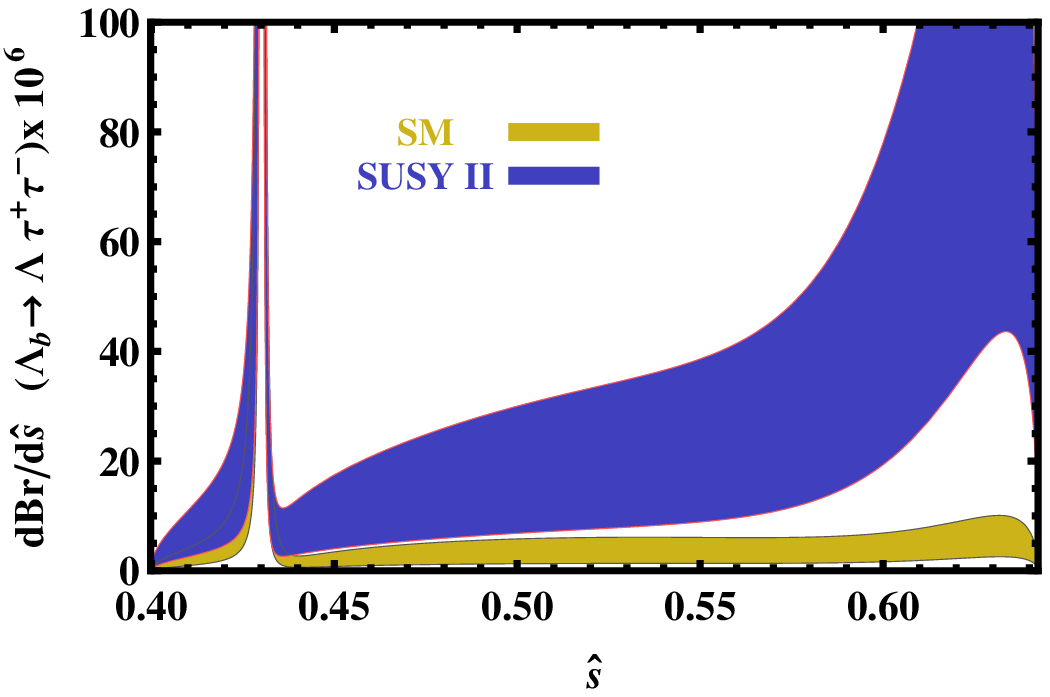,width=0.35\linewidth,clip=}
\end{tabular}
\begin{tabular}{ccc}
\epsfig{file=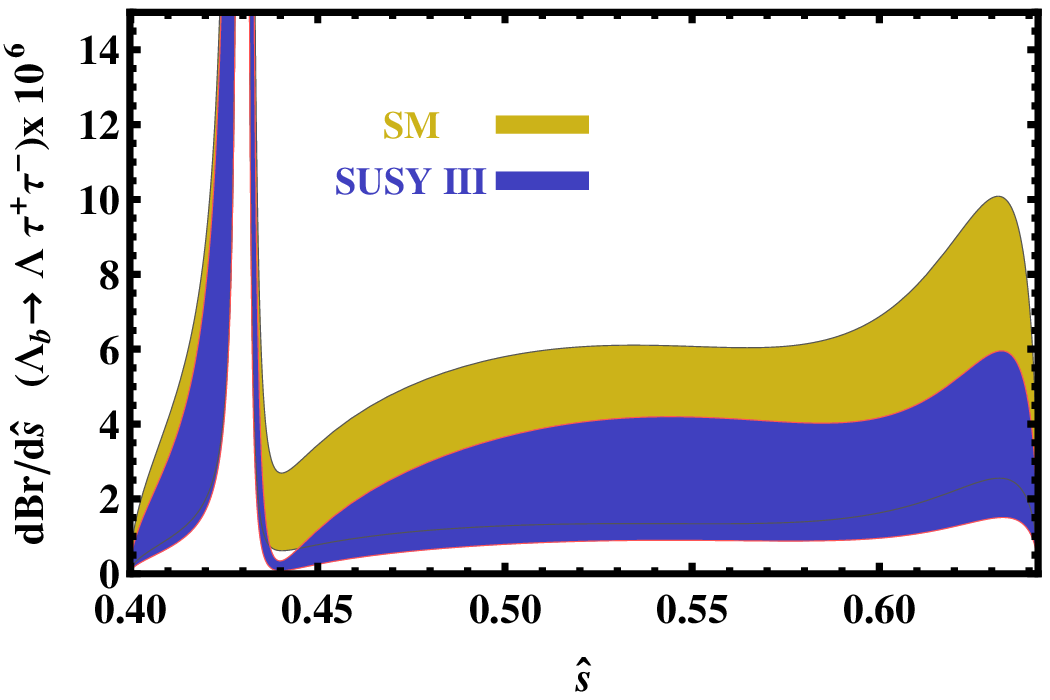,width=0.35\linewidth,clip=} &
\epsfig{file=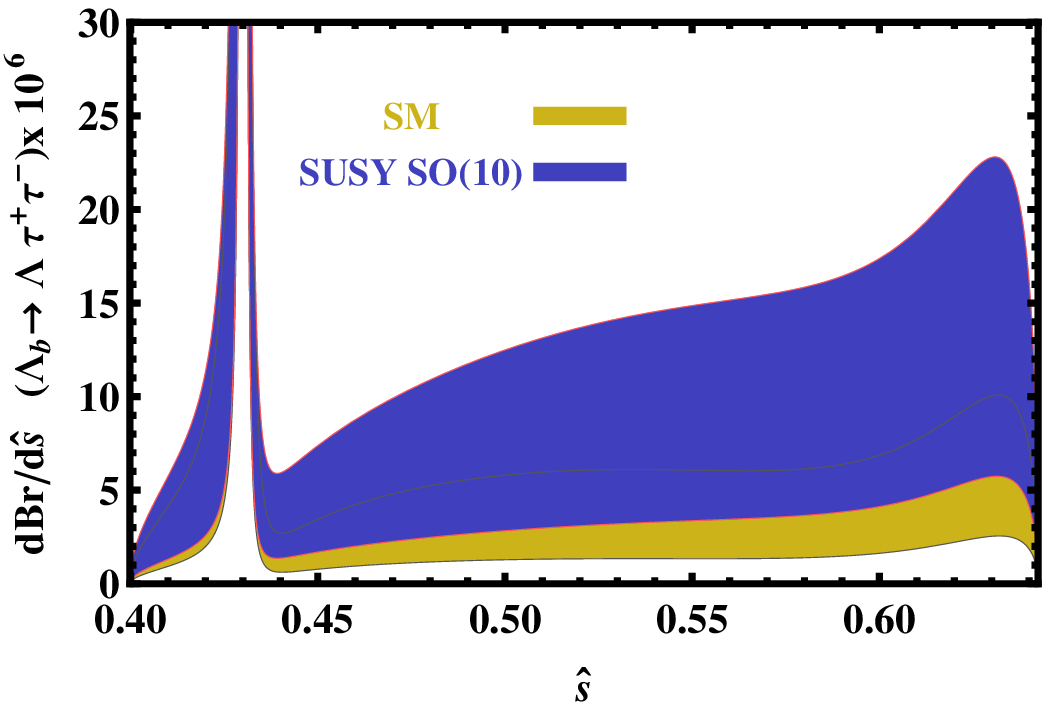,width=0.35\linewidth,clip=}
\end{tabular}
\caption{The same as Figure 5 but for $\tau$.}
\end{figure}
\begin{figure}[h!]
\centering
\begin{tabular}{ccc}
\epsfig{file=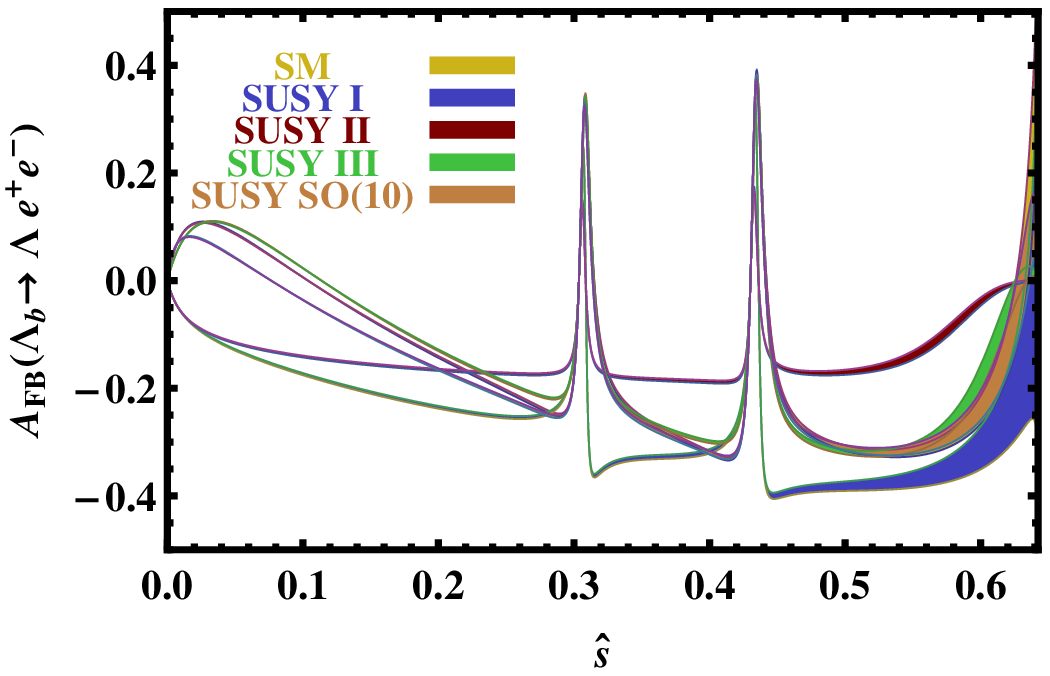,width=0.35\linewidth,clip=} &
\epsfig{file=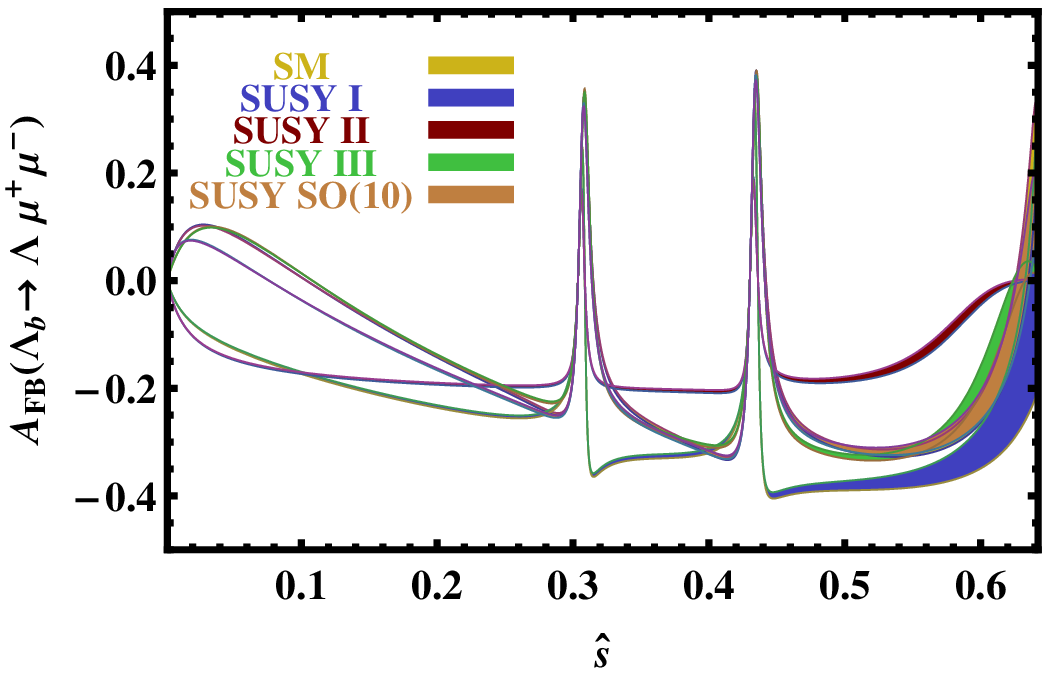,width=0.35\linewidth,clip=} 
\end{tabular}
\begin{tabular}{ccc}
\epsfig{file=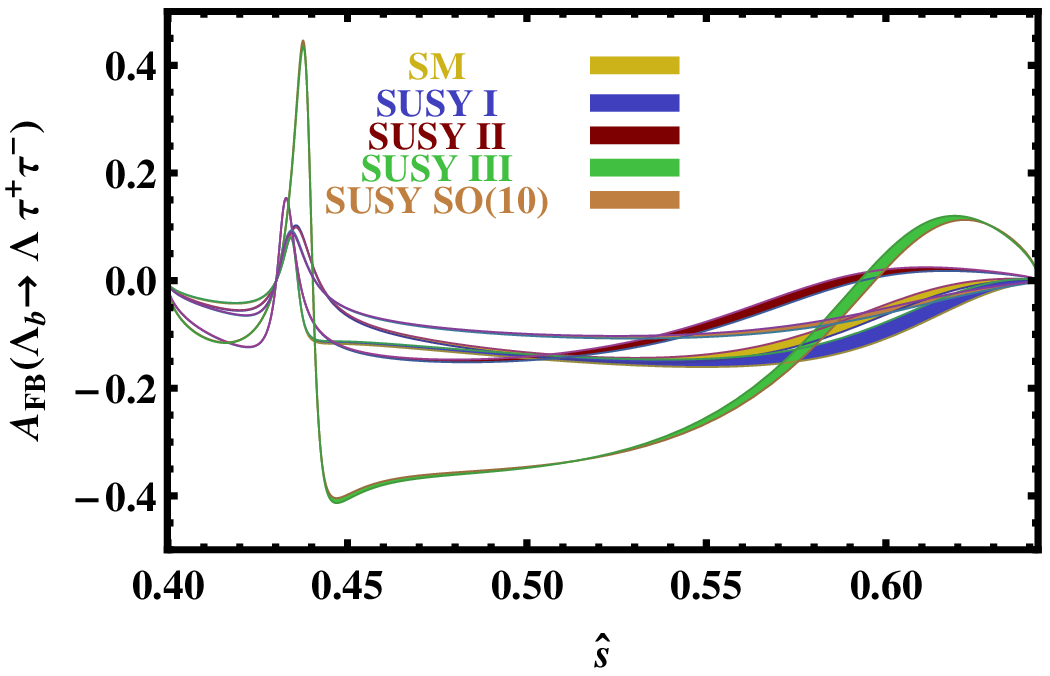,width=0.35\linewidth,clip=}
\end{tabular}
\caption{Comparison of the results of the FBA  with respect to  $\hat s$  for all lepton channels obtained  from different
 SUSY models with that of the   SM  considering the errors of form factors.}
\end{figure}
\begin{table}[ht]
\centering
\rowcolors{1}{lightgray}{white}
\begin{tabular}{cccc}
\hline \hline
 \mbox{\textit{BR}($\Lambda_{b}\rightarrow \Lambda e^{+} e^{-}$)} &\mbox{\textit{ Region I}}&\mbox{\textit{ Region II}}&\mbox{\textit{ Region III}}
                \\
\hline \hline
 SM                         & $(2.86 \pm 1.43)\times 10^{-6}$ & $(1.12 \pm 0.56)\times 10^{-6}$ &  $(0.81 \pm 0.40)\times 10^{-6}$  \\

 SUSY I                      & $(4.57 \pm 2.42)\times 10^{-6}$ &  $(1.56 \pm 0.82)\times 10^{-6}$ &  $(0.99 \pm 0.52)\times 10^{-6}$   \\

 SUSY II                     & $(5.66 \pm 3.05)\times 10^{-6}$ &  $(2.69 \pm 1.45)\times 10^{-6}$ & $(2.61 \pm 1.40)\times 10^{-6}$    \\

 SUSY III                    & $(2.93 \pm 1.52)\times 10^{-6}$ &  $(0.98 \pm 0.50)\times 10^{-6}$ &  $(0.66 \pm 0.34)\times 10^{-6}$  \\

 SUSY SO(10)                & $(2.65 \pm 1.35)\times 10^{-6}$ &  $(1.20 \pm 0.61)\times 10^{-6}$ &  $(1.09 \pm 0.55)\times 10^{-6}$   \\
 
\hline \hline
\end{tabular}
\caption{The values  of branching ratio for $\Lambda_{b}\rightarrow \Lambda e^{+} e^{-}$ decay channel at different regions in SM and different SUSY models considering the uncertainties of form factors.}
\end{table}
\begin{table}[ht]
\centering
\rowcolors{1}{lightgray}{white}
\begin{tabular}{cccc}
\hline \hline
\mbox{\textit{BR}($\Lambda_{b}\rightarrow \Lambda \mu^{+} \mu^{-}$)}& \mbox{\textit{ Region I}} & \mbox{\textit{ Region II}}& \mbox{\textit{ Region III}}
                  \\
\hline \hline
SM                         & $(2.25 \pm 1.12)\times 10^{-6}$ & $(1.12 \pm 0.56)\times 10^{-6}$ &  $(0.81 \pm0.40)\times 10^{-6}$  \\

 SUSY I                      & $(3.69 \pm1.95)\times 10^{-6}$ &  $(1.56\pm 0.82)\times 10^{-6}$ &  $(0.99 \pm0.52)\times 10^{-6}$   \\

 SUSY II                     & $(4.65 \pm2.51)\times 10^{-6}$ &  $(2.63 \pm 1.42)\times 10^{-6}$ & $(2.55\pm1.37)\times 10^{-6}$    \\

 SUSY III                    & $(2.04\pm1.06)\times 10^{-6}$ &  $(0.97 \pm 0.50)\times 10^{-6}$ &  $(0.65\pm0.33)\times 10^{-6}$  \\

 SUSY SO(10)                & $(2.33\pm1.18)\times 10^{-6}$ &  $(1.20\pm0.61)\times 10^{-6}$ &  $(1.09\pm0.55)\times 10^{-6}$  \\
\hline\hline
\end{tabular}
\caption{The values of branching ratio  for $\Lambda_{b}\rightarrow \Lambda \mu^{+} \mu^{-}$ decay channel at different regions in SM and different SUSY models considering the uncertainties of form factors.}
\end{table}
\begin{table}[ht]
\centering
\rowcolors{1}{lightgray}{white}
\begin{tabular}{ccc}
\hline \hline
 \mbox{\textit{BR}($\Lambda_{b}\rightarrow \Lambda \tau^{+} \tau^{-}$)} &\mbox{\textit{ Region I}}&\mbox{\textit{Region II }}
                \\
\hline \hline
 SM                         & $(0.87\pm0.43)\times 10^{-7}$ & $(3.84\pm1.92)\times 10^{-7}$   \\

 SUSY I                      & $(1.35\pm0.71)\times 10^{-7}$ &  $(5.55\pm2.94)\times 10^{-7}$  \\

 SUSY II                     & $(2.01\pm1.08)\times 10^{-7}$ &  $(2.44\pm1.31)\times 10^{-6}$  \\

 SUSY III                    & $(0.72\pm0.37)\times 10^{-7}$ &  $(2.13\pm1.10)\times 10^{-7}$  \\

 SUSY SO(10)                & $(1.12\pm0.57)\times 10^{-7}$ &  $(1.61\pm0.82)\times 10^{-6}$  \\
 
\hline \hline
\end{tabular}
\caption{The values of branching ratio  for $\Lambda_{b}\rightarrow \Lambda \tau^{+} \tau^{-}$ decay channel at different regions in  SM and different SUSY models considering the uncertainties of form factors.}
\end{table}
 In this subsection, we would like to consider the above mentioned physical quantities taking into account the uncertainties of the form factors and discuss the effects of these errors on the results. For this aim,
considering the errors of the form factors, we plot the dependence of the differential branching ratio and FBA on   $\hat s$ at different lepton channels and different models  in figures 5-8. From these figures we see
 that,
\begin{itemize}
 \item as far as the differential branching ratio are concerned, at the $e$ and $\mu$ channels, the band of SUSY SO(10) approximately covers the band of  the SM. In the case of  the SUSY I, II and III, although their bands coincide 
with that of the SM, there are some regions that these SUSY models have different predictions. Among these different  SUSY models, the maximum discrepancy from the SM prediction belongs to the SUSY II. In $\tau$ channel, the difference between
different SUSY models predictions and that of the SM can not be completely  killed by the errors of form factors for any SUSY models. In this channel, there are also common regions between the bands of  SUSY models and that of the SM, except than 
the SUSY II which does not approximately  coincide anywhere  with the SM result.
\item In the case of the FBA, the errors of form factors do not approximately affect the central values, except than the higher  values of  $\hat s$, which we see narrow bands for different SUSY models as well as the SM.
At $e$ and $\mu$ channels, the bands of the SUSY I, SUSY III,  SUSY SO(10) and SM coincide with each other somewhere at higher  values of  $\hat s$, but the SUSY II has different prediction. 
At $\tau$ channel, all models have different predictions.
\item As it is expected, the forward-backward asymmetry and in particular its zero-crossing points are more robust than the differential 
branching ratio such that they are not approximately   affected by the uncertainties of the form factors. This is the case also in $B \to K^{(*)} l^+ l^-$ channel.

Now, we discuss the effects of the uncertainties of the form factors for the branching ratios at different regions and for different models. Taking into account the errors of form factors, we present the values of 
branching ratios at different channels in Tables 9, 10 and 11. From these Tables we deduce the same results as we have seen from the figures of the differential branching ratio. Although the central
values for different SUSY models and the SM differ considerably from each other, considering the errors of 
the presented results in these Tables, we observe that approximately in all cases the results of different models coincide with each other, except for the SUSY II and SUSY SO(10) at $\tau$ channel and region II which
have considerable discrepancy with the other model predictions.
\end{itemize}

\section{Conclusion}
In the present work, we have calculated the amplitude and differential decay rate for the semileptonic  $\Lambda_b \rightarrow \Lambda \ell^+ \ell^-$ transition in different supersymmetric models.
We  have taken into account all twelve form factors entered the low energy matrix elements and recently calculated via light cone QCD sum rules in full theory to analyze the differential branching ratio, total branching fraction
and the lepton forward-backward asymmetry.
We have considered different SUSY scenarios in the calculations and compared the obtained results with the SM predictions. As far as the central values of the form factors are considered, in general, 
we observed  considerable deviations from the SM predictions. In the case of the (differential) branching ratio, the maximum deviations from the SM predictions belong to the SUSY II scenario. 
As far as the FBA is concerned,
at $e$ and $\mu$ channels and lower values of the $\hat s$, the  maximum deviation
belongs to the SUSY I, however, for the  higher values of the  $\hat s$ and the same lepton channels, the maximum discrepancy corresponds again  to the SUSY II model.
Taking into account the uncertainties of the form factors, we have observed that the branching ratio is more affected by these errors. The bands of the SUSY SO(10) approximately cover the SM bands at the $e$ and $\mu$
channels. For other SUSY models and all lepton channels, although we have seen some intersection regions between different SUSY bands and the SM predictions, there are considerable discrepancies between the SM and SUSY 
models predictions. Especially, at $\tau$ channel, there is a big discrepancy between the SUSY II and the SM bands. When we consider the FBA, the uncertainties of the form factors do not affect this
 quantity and  its zero-crossing points. We see overall a considerable discrepancies between the narrow bands of the different considered SUSY models and that of the SM. Such discrepancies
  can be considered
as a signal for existence of  the supersymmetric particles.

The orders of the branching ratio at all lepton channels and all the considered regions of $q^2$ depict that these decay channels can be checked at LHC in near future. Note that as we have also previously stressed,
this channel for $\mu$ case has been observed recently by CDF Collaboration at Fermilab. We are waiting for the LHCb Collaboration results on these channels, which they have in their
 physics program \cite{LHCb}.

Comparison of the experimental results on the branching ratio as well as the FBA with
the predictions of the present work, especially determination of the sign and zero-crossing points of the FBA, which have not been affected by the errors of the form factors,  can help us get valuable information about the existence of the SUSY particles.  

As we have already noticed, in numerical analysis, we have used the  values of the  Wilson coefficients presented in Table 5 for   different SUSY scenarios. These values are obtained when the 
masses of the  neutral Higgs bosons are taken in the interval $(91-200)~GeV$ (see for instance \cite{Wil.coef,Higgs,Higgs2,Higgs3}). Considering the recent developments by the CMS and ATLAS Collaborations at CERN on the 
mass of the Higgs-like boson ($\sim 125~GeV$), the used values of the Wilson coefficients are still viable. However, after clarifying   whether the obtained boson at LHC  is the standard or non-standard Higgs, 
and using its exact mass, one may recalculate the Wilson coefficients in different SUSY scenarios.
Obviously, it will be possible to improve the obtained results in the present work using the new values of the Wilson coefficients.

                                 
                            
\end{document}